# On the Electron Pairing Mechanism of Copper-Oxide High Temperature Superconductivity

Shane M. O'Mahony[1,§], Wangping Ren[2,§], Weijiong Chen[2], Yi Xue Chong[3], Xiaolong Liu[3,4], H. Eisaki[5], S. Uchida[6], M.H. Hamidian[3], and J.C. Séamus Davis[1,2,3,7]

1. *Department of Physics, University College Cork, Cork T12 R5C, Ireland*
2. *Clarendon Laboratory, University of Oxford, Oxford, OX1 3PU, UK.*
3. *Department of Physics, Cornell University, Ithaca NY 14850, USA*
4. *Kavli Institute for Nanoscale Science, Cornell University, Ithaca, NY 14853, USA*
5. *Inst. of Advanced Industrial Science and Tech., Tsukuba, Ibaraki 305-8568, Japan.*
6. *Department of Physics, University of Tokyo, Bunkyo, Tokyo 113-0011, Japan.*
7. *Max-Planck Institute for Chemical Physics of Solids, D-01187 Dresden, Germany*
§ *These authors contributed equally to this project.*

The elementary $CuO_2$ plane sustaining cuprate high temperature superconductivity occurs typically at the base of a periodic array of edge-sharing $CuO_5$ pyramids. Virtual transitions of electrons between adjacent planar Cu and O atoms, occurring at a rate $t/\hbar$ and across the charge-transfer energy gap $\mathcal{E}$, generate 'superexchange' spin-spin interactions of energy $J \approx 4t^4/\mathcal{E}^3$ in an antiferromagnetic correlated-insulator state. However, hole doping this $CuO_2$ plane converts this into a very high temperature superconducting state whose electron-pairing is exceptional. A leading proposal for the mechanism of this intense electron-pairing is that, while hole doping destroys magnetic order it preserves pair-forming superexchange interactions governed by the charge-transfer energy scale $\mathcal{E}$. To explore this hypothesis directly at atomic-scale, we combine single-electron and electron-pair (Josephson) scanning tunneling microscopy to visualize the interplay of $\mathcal{E}$ and the electron-pair density $n_P$ in $Bi_2Sr_2CaCu_2O_{8+x}$. The responses of both $\mathcal{E}$ and $n_P$ to alterations in the distance $\delta$ between planar Cu and apical O atoms are then determined. These data reveal the empirical crux of strongly correlated superconductivity in $CuO_2$, the response of the electron-pair condensate to varying the charge transfer energy. Concurrence of predictions from strong-correlation theory for hole-doped charge-transfer insulators with these observations, indicates that charge-transfer superexchange is the electron-pairing mechanism of superconductive $Bi_2Sr_2CaCu_2O_{8+x}$.

***Concept of Electron-pairing from Charge-transfer Superexchange Interactions***

***1*** The prospect that hole-doped $CuO_2$ retains charge-transfer superexchange interactions between adjacent Cu spins has long motivated an hypothesis that spin-singlet electron-pair formation mediated by superexchange, is the mechanism of high temperature superconductivity. In transition-metal oxide insulators, superexchange (*1*) generates intense magnetic interactions between electrons that are localized at adjacent transition-metal atoms, typically generating robust antiferromagnetism. The superexchange interaction *J* occurs when the degeneracy of transition-metal *3d* orbitals is lifted by the Coulomb energy *U* required for their double occupancy, so that intervening oxygen *2p* energy levels are separated from the relevant transition-metal *3d* level by the charge-transfer energy $\mathcal{E}$. Within the framework of the three-band model, the interactions of two adjacent *3d* electrons of spin $\boldsymbol{S}_i$ are well approximated by a Heisenberg Hamiltonian $H = J\boldsymbol{S}_i \cdot \boldsymbol{S}_j$, with *J* the superexchange interaction produced by a multi-stage process of electronic exchange between spins on adjacent 3d orbitals via the non-magnetic oxygen *2p* orbitals. In the strong-coupling limit $U/t \gg 1$, $J \approx 4t^4/\mathcal{E}^3$ where the transition rate of electrons between *3d* and *2p* orbitals is given by $t/\hbar$. Specifically for $CuO_2$-based materials, the planar $Cu^{2+}$ ions are in the *3d⁹* configuration with a singly occupied $d_{x^2-y^2}$ orbital, while the planar $O^{2-}$ ions have closed *2p⁶* shells whose in-plane $p_\sigma$ orbitals dominate. To doubly occupy any $d_{x^2-y^2}$ orbital requires an energy *U* so great that the *d*-electrons become fully Mott-localized in a charge-transfer insulator state, with the $p_\sigma$ energy level separated from the pertinent $d_{x^2-y^2}$ level by the $CuO_2$ charge-transfer energy $\mathcal{E}$ (Fig. 1a). Under such circumstances, an electronic structure with $t \approx 0.4$ eV and $\mathcal{E} \approx 1$ eV implies a superexchange energy $J \approx 100$ meV that should stabilize a robust spin-1/2, $\boldsymbol{Q} = (\pi, \pi)$ antiferromagnetic state (Fig. 1b). Just such a state is observed (*2*), confirming that charge-transfer superexchange is definitely the mechanism of the $CuO_2$ antiferromagnetic state. However, when holes are doped into the $CuO_2$-plane they enter the $p_\sigma$ orbitals, both disrupting the antiferromagnetic order and delocalizing the electrons. This situation may be approximated using the three-band Hamiltonian based on a single Cu $d_{x^2-y^2}$ plus two O $p_\sigma$ orbitals per unit cell (*3, 4*):

$$H = \sum_{i\alpha j\beta,\sigma} t_{ij}^{\alpha\beta} c_{i\sigma}^{\dagger\alpha} c_{j\sigma}^{\beta} + \sum_{i\alpha,\sigma} \varepsilon_\alpha n_{i\sigma}^\alpha + U \sum_i n_{i\uparrow}^d n_{i\downarrow}^d \qquad (1)$$

Here *i,j* enumerate planar $CuO_2$ unit cells, $\alpha,\beta$ label any of the three orbitals, $t_{ij}^{\alpha\beta}$ are transition rates for electrons between orbitals $\alpha,\beta$ at sites *i, j*, $\varepsilon_\alpha$ are the orbital energies, and $n_{i\uparrow}^d, n_{i\downarrow}^d$ are the $d_{x^2-y^2}$ orbital occupancies by spin state. Heuristically, such models describe a two-dimensional correlated metallic state with intense antiferromagnetic spin-spin interactions. If superconductivity occurs (Fig. 1c), it is signified by the appearance of a condensate of electron-pairs $\Psi \equiv \langle c_{i\downarrow}^d c_{j\uparrow}^d \rangle$, a phenomenon that is now directly accessible to visualization using scanned Josephson tunneling microscopy (SJTM) (*5, 6, 7, 8, 9*).

**2**     Empirical study of charge-transfer superexchange as the mechanism of this superconductivity requires knowledge of the dependence of $\Psi$ on the charge-transfer energy $\mathcal{E}$, but this has not been experimentally accessible. Certainly, $\mathcal{E}$ and $J$ have long been studied using optical reflectivity, Raman spectroscopy, tunneling spectroscopy, angle resolved photoemission, and resonant inelastic X-ray scattering (Ref. *10*, Section I). Typically, to access different $\mathcal{E}$ for these studies required changing between crystal families in the antiferromagnetic-insulator state. But this renders impossible the required comparison between $\mathcal{E}$ and $\Psi$ measured simultaneously in the same superconductive state. Instead, the maximum superconducting critical temperature $T_C$ subsequent to hole-doping is often proposed as a proxy for $\Psi$ and then compared with the $\mathcal{E}$ derived from the parent insulator, for a range of different compounds. But varying the crystal family alters a wide variety of other material parameters besides $\mathcal{E}$, and $T_C$ is anyway controlled by other influences including dimensionality and superfluid phase-stiffness (*11*). More fundamentally, advanced theoretical analysis has recently revealed that no one-to-one correspondence exists between the $T_C$ and $\Psi$ in the CuO₂ Hubbard model (*12, 13*). Hence, although greatly encouraging, studies comparing maximum superconductive $T_C$ with insulating $\mathcal{E}$, cannot be conclusive as to the electron-pairing mechanism. On the other hand, muon spin rotation studies do make clear that $\Psi$ diminishes rapidly with increasing correlations upon approach the charge-transfer insulator state (*14*). Ultimately, to identify the essential physics subtending this electron pairing, a direct and systematic measurement of the dependence of the electron-pair condensate $\Psi$ on the charge-transfer energy $\mathcal{E}$ at the same hole density, is required.

**3**     In this context, dynamical mean-field theory (DMFT) analysis of the CuO₂ Hubbard model has recently yielded quantitative predictions of how $\Psi$ is controlled by $\mathcal{E}$.

Moreover, theory also indicates that this interplay may be adjusted by altering the distance $\delta$ between each Cu atom and the apical O atom of its $CuO_5$ pyramid (*15, 16, 17, 18*). This is because varying $\delta$ should alter the Coulomb potential at the planar Cu and O atoms, modifying $\mathcal{E}$ and thereby controlling $\Psi$ in a predictable manner (*16, 17, 18*), a scenario which has been advocated since the discovery of cuprate superconductivity (*19, 20, 21, 22*). These realistically parameterized, quantitative predictions represent an exciting new opportunity: measurement of the dependences of $\Psi$ on $\mathcal{E}$ at the Cu atom beneath each displaced apical oxygen atom, potentially yielding quantitative knowledge of $d\Psi/d\mathcal{E}$ as a direct test for a charge-transfer superexchange electron-pairing mechanism (*16,17,18*). For experimentalists, the challenge is thus to measure the relationship between $\Psi$ and $\mathcal{E}$ directly and simultaneously at the superconducting $CuO_2$-plane. If available, such data could play a role analogous to the *isotope effect* in conventional superconductors (*23*), by identifying empirically for cuprates the specific electron-electron interaction that controls electron pair formation.

### Techniques for Visualization of Charge-transfer Energy and Electron Pair Density

**4**       To explore this prospect, one must measure $\Psi$ and $\mathcal{E}$ as a function of separation $\delta$ above each planar Cu atom. But $\Psi$ is, in general, a complex-valued field and thus not a physical observable, meaning that experimentalists must study $|\Psi|^2 \equiv n_P$, the electron-pair density. Moreover, the pseudogap masks the true electron-pairing energy gap so that single-particle tunneling spectroscopy cannot be used to image the superconductive order parameter in lightly hole-doped cuprates. Our strategy therefore combines novel techniques in atomic resolution imaging with a fortuitous property of the canonical cuprate $Bi_2Sr_2CaCu_2O_{8+x}$. First, a mismatch between preferred bond lengths of the rock-salt and perovskite layers in $Bi_2Sr_2CaCu_2O_{8+x}$ generates a $\lambda \sim 26$Å periodic modulation of unit-cell dimensions (Fig. 1d), along the crystal *a*-axis or equivalently the (1,1) axis of the $CuO_2$ plane (*24*). Providentially, this *crystal supermodulation* generates periodic variations in $\delta$ by up to 12%, in the single-electron excitation spectrum (*25*), and in the electron-pair (Josephson) current (*7*). However, the influence of the supermodulation on $\mathcal{E}$ and $n_P$ were unknown. Crucially for our objectives, the value of $\delta$ at every location $\boldsymbol{r}$ can be evaluated by atomic-resolution imaging of the supermodulation in topographic images $T(\boldsymbol{r})$ measured at the crystal's BiO termination layer (Fig 1d, Fig. 2a), and then by using X-ray crystallography to relate $T(\boldsymbol{r})$ to the spatial pattern of apical displacements $\delta(\boldsymbol{r})$ just underneath (Ref. 10 Section II). Second, by measuring differential tunnel conductance $dI/dV(\boldsymbol{r},V) \equiv g(\boldsymbol{r},V)$ as a function of location $\boldsymbol{r}$ and tip-sample voltage V, the density of electronic states $N(\boldsymbol{r},E) \propto g(\boldsymbol{r},V=E/e)$ can be visualized for the high energy range governed by Eqn. 1. In principle, this allows energy scales such as $\mathcal{E}(\boldsymbol{r})$ in the spectrum of $Bi_2Sr_2CaCu_2O_{8+x}$ to be determined versus location $\boldsymbol{r}$. Third, using superconducting STM tips ($Bi_2Sr_2CaCu_2O_{8+x}$ nanoflake tips (7)) to image the Josephson critical current $I_J$ for electron-pair tunneling versus location $\boldsymbol{r}$, allows direct visualization of sample's electron-pair density (7,8,9) $n_P(\boldsymbol{r}) \equiv |\Psi|^2 \propto \left(I_J(\boldsymbol{r})R_N(\boldsymbol{r})\right)^2$ where $R_N$ is the tip-sample normal state junction resistance. Thus, our concept is to visualize both $\mathcal{E}(\boldsymbol{r})$ and $n_P(\boldsymbol{r})$ directly at atomic scale, as a function of the apical oxygen displacements $\delta(\boldsymbol{r})$ that are produced by the crystal supermodulation in $Bi_2Sr_2CaCu_2O_{8+x}$.

**5**     In practice, single crystals of $Bi_2Sr_2CaCu_2O_{8+x}$ with hole-density $p \approx 0.17$ are cleaved in cryogenic ultrahigh vacuum in a dilution refrigerator based spectroscopic imaging scanning tunnelling microscope (SISTM), to reveal the BiO termination layer (Fig. 2a). The $CuO_2$ plane is $\sim 5\text{Å}$ beneath the BiO surface and separated from it by the SrO layer containing the apical oxygen atom of each $CuO_5$ pyramid (Fig. 1a). A surface corrugation $T(\boldsymbol{r}) = A(\boldsymbol{r})\cos\Phi(\boldsymbol{r})$, where $\Phi(\boldsymbol{r}) = \boldsymbol{Q_S} \cdot \boldsymbol{r} + \theta(\boldsymbol{r})$, occurs at the bulk supermodulation wavevector $\boldsymbol{Q_S} \cong (0.15,0.15)2\pi/a_0$ where $\theta(\boldsymbol{r})$ describes effects of disorder (Fig. 2a). The supermodulation phase $\Phi(\boldsymbol{r})$ is then imaged by analyzing $T(\boldsymbol{q})$, the Fourier transform of $T(\boldsymbol{r})$, with typical results shown in Fig. 2b (Ref. 10 Section II). X-ray scattering studies of the $Bi_2Sr_2CaCu_2O_{8+x}$ crystal supermodulation demonstrate that the distance to apical oxygen atom $\delta$ is minimal at $\Phi = 0$ and maximal at $\Phi = \pi$, because the displacement amplitude of the $c$-axis supermodulation is greater in the $CuO_2$ layer than in the adjacent SrO layer. Thus, $\delta(\boldsymbol{r})$ is determined from the measured $\Phi(\boldsymbol{r})$ based on X-ray refinement as: $\delta(\Phi) \approx 2.44 - 0.14\cos(\Phi)$ Å (Ref. *10*, Section II). For example, the apical displacement imaging results $\delta(\boldsymbol{r})$ from Fig. 2a,b are shown in Fig. 2c. This same $\Phi(\boldsymbol{r}):\delta(\boldsymbol{r})$ procedure is used throughout our study.

***Coterminous Visualization of Charge-transfer Energy and Electron Pair Density***
**6**     In search of associated modulations in $\mathcal{E}(\boldsymbol{r})$, Fig. 3a shows a typical topographic image of the BiO termination layer, while Fig. 3b shows two high-voltage single-electron $g(V)$ spectra measured using junction-resistance $R_N \approx 85$ GΩ in the same field of view. Such enormous junction resistances (or large tip-sample distances) preclude effects on $g(V)$ of the tip-sample electric field. Hence, by visualizing $g(\boldsymbol{r}, V)$ in the $-1.6$ V $\leq V \leq 2$ V range at these junction-resistances, one can determine empirically if $\mathcal{E}(\boldsymbol{r})$ modulations exist. For example, Fig. 3b shows representative $g(\boldsymbol{r}, V)$ spectra plotted on a logarithmic scale. We use the standard approach to estimate $\mathcal{E}$ as being the minimum energy difference between upper and lower bands (*26*) at a constant conductance $G \approx 20$ pS, as shown by double headed arrows. This value of $G$ implies no overlap in the measurements of $\mathcal{E}$ with the range of voltages $|V| > 0.9\ V$ where oxygen dopant atoms or vacancies cause significant disorder as indicated in SM Fig, 4 of Ref. 10. Thus, the minimum energy separation between the top of the lower band and bottom of the upper band is indicated by the horizontal double headed arrows (Ref. 10 Section III). The blue arrows represent $\mathcal{E}(\Phi = 0)$ and the red arrows $\mathcal{E}(\Phi = \pi)$. This is consistent with the well-established (*27, 28, 29*) value of charge-transfer energy $\mathcal{E} \approx 1.2$eV in $Bi_2Sr_2CaCu_2O_8$ (approximated by grey shaded region in Fig. 3b). Finally, plotting $g(\boldsymbol{r}, V)$ in Fig. 3c, along the trajectory shown by the dashed line Fig. 3a, reveals directly that $\mathcal{E}(\boldsymbol{r})$ modulates strongly at the supermodulation wavevector, with $\mathcal{E}(\Phi = 0) \approx 1.35$eV (blue arrows) and $\mathcal{E}(\Phi = \pi) \approx 0.95$eV (red arrows), as indicated.

**7**      Correspondingly, to search for modulations in $n_P(\boldsymbol{r})$, Fig. 3d shows a typical topographic image of the BiO termination layer using a tip terminating in a $Bi_2Sr_2CaCu_2O_{8+x}$ nanoparticle (*7*). The junction resistance used here is $R_N \approx 21$ MΩ; this is almost five thousand times lower than that used for the $\mathcal{E}(\boldsymbol{r})$ studies, as are the typical electron-pair tunneling voltages $V_J$. Figure 3e shows a typical $I_P(V_J)$ spectrum measured in this field of view, with the tip-sample Josephson junction exhibiting a phase-diffusive steady-state at voltage $V_J$, with electron-pair current $I_P(V_J) = \frac{1}{2}I_J^2 Z V_J / (V_J^2 + V_c^2)$ where $Z$ the high-frequency junction impedance and $V_C$ is the voltage for maximum $I_P(V_J)$. Then, because the maxima in $I_P(V_J)$ occur at $I_m \propto I_J^2$, atomic-scale visualization of an electron-pair density is achieved (*7, 8, 9*) as $n_P(\boldsymbol{r}) \propto I_m(\boldsymbol{r})R_N^2(\boldsymbol{r})$ or equivalently $n_P(\boldsymbol{r}) \propto g_0(\boldsymbol{r})R_N^2(\boldsymbol{r})$ (Ref. 10 Section IV). In this study, we use the protocol $n_P(\boldsymbol{r}) \propto g_0(\boldsymbol{r})R_N^2(\boldsymbol{r})$ to produce all key quantitative results as presented in Figs. 4 and 5. However, one can visualize empirically if $n_P(\boldsymbol{r})$ modulations exist, by measuring $I_P(\boldsymbol{r}, V_J)$ along the trajectory of the supermodulation (dashed line Fig. 3d). The result, as shown in Fig. 3f, clearly demonstrates how $|I_m|$ also modulates strongly at wavevector $\boldsymbol{Q_S}$ (*7*).

**8**     Together, these data reveal that both the band-separation energy $\mathcal{E}(\boldsymbol{r})$ and the condensate electron-pair density $n_P(\boldsymbol{r})$ are modulated periodically, by the crystal supermodulation of Bi$_2$Sr$_2$CaCu$_2$O$_{8+x}$. To quantify and relate these phenomena, we consider two exemplary fields of view whose $T(\boldsymbol{r})$ are shown Fig. 4a,b. Both $T(\boldsymbol{r})$ images are evaluated to determine their separate $\Phi(\boldsymbol{r})$, with the ends of the $\Phi = \pi$ contours indicated by the arrowheads in each. A high-voltage single-electron tunnelling $g(\boldsymbol{r}, V)$ map is measured at $R_N \approx 85$ G$\Omega$ and $T = 4.2$ K in the field of view (FOV) of Fig. 4a, while a low-voltage electron-pair tunnelling $I_P(V_J)$ map at $R_N \approx 21$ M$\Omega$ and $T = 2$ K is measured in that of Fig. 4b. To visualize $\mathcal{E}(\boldsymbol{r})$, we estimate $\mathcal{E}$ to be the minimum energy difference between upper and lower bands (*26*) at a constant conductance $G \approx 20$ pS. The resulting $\mathcal{E}(\boldsymbol{r})$ shown in Fig. 4c is correctly representative and appears little different if we estimate $\mathcal{E}(\boldsymbol{r})$ anywhere in the range $20\text{pS} \leq G \leq 80$ pS (Ref. 10 Section III). Concomitantly, to visualize $n_P(\boldsymbol{r})$ we measure $g_0(\boldsymbol{r})$ and multiply by the measured $R_N^2(\boldsymbol{r})$ modulations from the same FOV as Fig. 4b. The normal-state junction resistance $R_N(\boldsymbol{r})$ is obtained by self-normalizing two sets of $dI/dV(\boldsymbol{r})$ spectra, one for $V_{max} < \Delta/e$ and the other for $V_{max} > \Delta/e$ , measured in precisely the same FOV (Ref. *10*, Section IV). Thus, Fig. 4d shows measured $n_P(\boldsymbol{r})$ in the FOV of Fig. 4b. Finally, when Fig. 4c is Fourier-filtered at $\boldsymbol{Q_S}$ it reveals the first-harmonic modulations in $\tilde{\mathcal{E}}(\boldsymbol{r})$ as presented in Fig. 4e, while identical filtering of Fig. 4d at $\boldsymbol{Q_S}$ yields the first-harmonic modulations in $\tilde{n}_P(\boldsymbol{r})$ as seen in Fig. 4f. Thus, visualization of the crystal supermodulation effect on both $\mathcal{E}(\boldsymbol{r})$ and $n_P(\boldsymbol{r})$, simultaneously with their $\Phi(\boldsymbol{r})$, is now possible in Bi$_2$Sr$_2$CaCu$_2$O$_{8+x}$.

*Synthesis*

**9**     So, how does supermodulation displacement of the apical oxygen atom $\delta(\boldsymbol{r})$ (and to a lesser extent that of other atoms) alter the charge transfer energy $\mathcal{E}(\boldsymbol{r})$ and the electron-pair density $n_P(\boldsymbol{r})$ at each planar Cu atom (*16-22*) in Bi$_2$Sr$_2$CaCu$_2$O$_{8+x}$? To synthesize data as in Fig. 4, we first plot apical distance alterations versus phase $\delta(\Phi)$ for Bi$_2$Sr$_2$CaCu$_2$O$_{8+x}$, as shown by grey dots in Fig. 5b. We then process $\mathcal{E}(\boldsymbol{r})$ retaining only wavevectors close to $\pm\boldsymbol{Q_S}$. Then, by corresponding simultaneous $\Phi(\boldsymbol{r})$ : $\mathcal{E}(\boldsymbol{r})$ measurements (e.g. Fig. 4a,c) we determine $\bar{\mathcal{E}}(\Phi)$ whose value is normalised to the mean measured value and shown as red dots in Fig. 5b; this is found to be a very repeatable characteristic of Bi₂Sr₂CaCu₂O₈₊ₓ. Similarly, by corresponding simultaneous $\Phi(r) : n_P(r)$ measurements (e.g. Fig. 4b,d) we determine $\bar{n}_P(\Phi)$ which is normalized to the mean value of measured $n_P(r)$. This is shown by blue dots in Fig. 5b; this is another repeatable characteristic (Ref. 10 Section VI). To maximize the precision of both the Fourier filtering and lock-in methods, we perform this analysis in an FOV which includes as many periods of the supermodulation as possible (for $\mathcal{E}(r)$ 7 periods and for $n_P(r)$ 13 periods). The microscopic relationship of $\mathcal{E}$ to $\delta$ can then be determined by eliminating common variable $\Phi$ from Fig. 5b. The result, shown in Fig. 5c, provides a direct measurement of this long-sought characteristic (*16-21*) of cuprate electronic structure: $d\mathcal{E}/d\delta \approx -1.04 \pm 0.12$ eV/Å and $d\bar{n}_P/d\delta \approx 0.85 \pm 0.22$ Å⁻¹ for Bi₂Sr₂CaCu₂O₈₊ₓ. More fundamentally, the atomic-scale relationship between the normalized electron-pair density $\bar{n}_P$ and the charge-transfer energy $\mathcal{E}$ is derived by eliminating the common variable $\Phi$. The result as shown in Fig. 5d, demonstrates that $d\bar{n}_P/d\mathcal{E} \approx -0.81 \pm 0.17$ eV⁻¹ or equivalently that $d\frac{|\langle c_\uparrow c_\downarrow \rangle|}{d\mathcal{E}} \approx -0.40 \pm 0.09$ eV⁻¹ over a wide range of charge-transfer energy scales in Bi₂Sr₂CaCu₂O₈₊ₓ.

**10**    Although the original predictions (*16, 17*) for $d\mathcal{E}/d\delta$ were for La₂CuO₄, they are still in reasonable agreement with our observations for Bi₂Sr₂CaCuO₈₊ₓ as shown in Fig. 5c. Theoretical  predictions for the direct effect on the cuprate electron-pair condensate of altering the charge-transfer $\mathcal{E}$ yield (Ref. 10 Section VII) $d|\overline{\langle c_\uparrow c_\downarrow \rangle}|/d\mathcal{E} \approx -\alpha/2$ eV⁻¹ or equivalently $d\bar{n}_P/d\mathcal{E} \approx -\alpha$ eV⁻¹, with a range $0.3 \lesssim \alpha \lesssim 1.0$ depending on the material-specific parameters (*12,16, 17, 18*). The precise parameters used in these calculations for a variety of different materials are given in Ref. (*16*). Figure 5d indicates the anticipated range of $\alpha$ for different materials using a yellow shaded triangle. For Bi₂Sr₂CaCu₂O₈ specifically (*12*), the three-band CuO₂ Hubbard model prediction for a superexchange electron-pairing mechanism is that $d|\overline{\langle c_\uparrow c_\downarrow \rangle}|/d\mathcal{E} \approx -0.46 \pm 0.05$ eV⁻¹ or equivalently that $\alpha \approx 0.93 \pm 0.1$ eV⁻¹. The agreement with experimental observations reported in Fig. 5d is self-evident.

**11**    For decades the electron-pairing mechanism of cuprate high-temperature superconductivity has been hypothesized (*30-37*) as due to electron-electron interactions mediated by superexchange, but with the electron-pair condensate $\Psi$ subject to the strong no-double-occupancy constraints on the Cu $d_{x^2-y^2}$ orbitals (*38, 39*), (Fig. 1c). When such interactions and constraints were studied using mean-field Gutzwiller projection (*31,32*), or by slave-boson techniques (*33,34,35*), or by Monte Carlo numerical techniques (*36,37*) the phase diagram and many other key characteristics that emerge are congruent with observations (*38,39*). Contemporary theoretical studies, using a wide variety of advanced theoretical and numerical techniques (*40-45*), also predict with growing confidence that it is the superexchange interaction which creates electron pairing in the three-band $CuO_2$ Hubbard model. However, direct experimental tests of the relationship between the cuprate electron-pair condensate and the charge transfer energy of this model, were non-existent. Here, by visualizing the electron-pair density $n_P(\boldsymbol{r})$ using SJTM (e.g. Fig. 4d,f), and the charge transfer energy $\mathcal{E}(\boldsymbol{r})$ using high-voltage SISTM (e.g. Fig. 4c,e), we find empirically that both modulate together at the $Bi_2Sr_2CaCu_2O_{8+x}$ crystal supermodulation wavevector $\boldsymbol{Q}s$ (Fig. 2b,c; Fig. 5b). This joint $\mathcal{E}(\boldsymbol{r}) : n_P(\boldsymbol{r})$ modulation is observed comprehensively throughout these studies of $Bi_2Sr_2CaCu_2O_8$, with its existence being independent of exactly which atomic displacements occur within the crystal supermodulation. The consequent demonstration that $d|\langle c_\uparrow c_\downarrow\rangle|/d\mathcal{E} < 0$ (Fig. 5d) is a direct visualization of an effect long anticipated in the theory of superexchange mediated electron pairing in cuprates (*3,4,15-18,30-39*), and from experiments based on muon spin rotation (*14*). More specifically, recent numerical studies of the three-band $CuO_2$ Hubbard model (*12*) within which charge-transfer superexchange is demonstrably the cause of electron pairing (*40-45*), yield quantitative agreement between predicted $d|\overline{\langle c_\uparrow c_\downarrow\rangle}|/d\mathcal{E} \approx -0.46 \pm 0.05$ $eV^{-1}$ and our experimental determination that $d\bar{n}_P/d\mathcal{E} \approx -0.81 \pm 0.17$ $eV^{-1}$ for $Bi_2Sr_2CaCu_2O_{8+x}$. Taken at face value, the data in Fig. 5 thus indicate that charge-transfer superexchange is key to the electron-pairing mechanism of the hole-doped cuprate superconductor $Bi_2Sr_2CaCu_2O_{8+x}$.

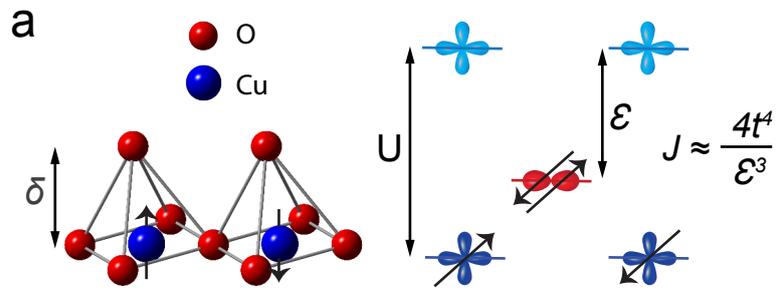

a

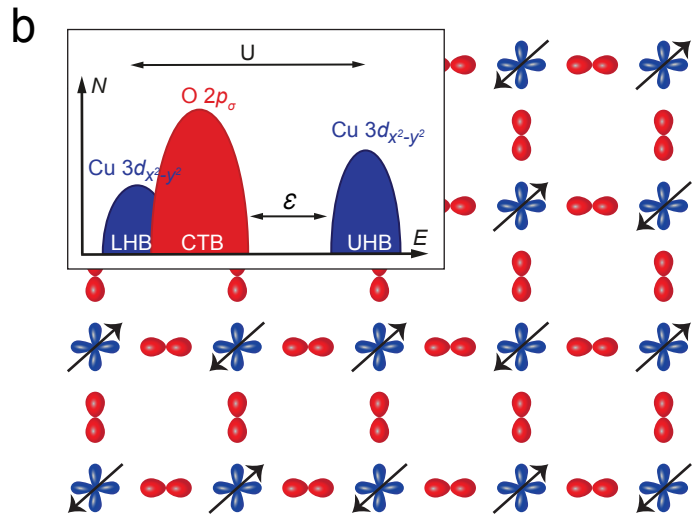

b

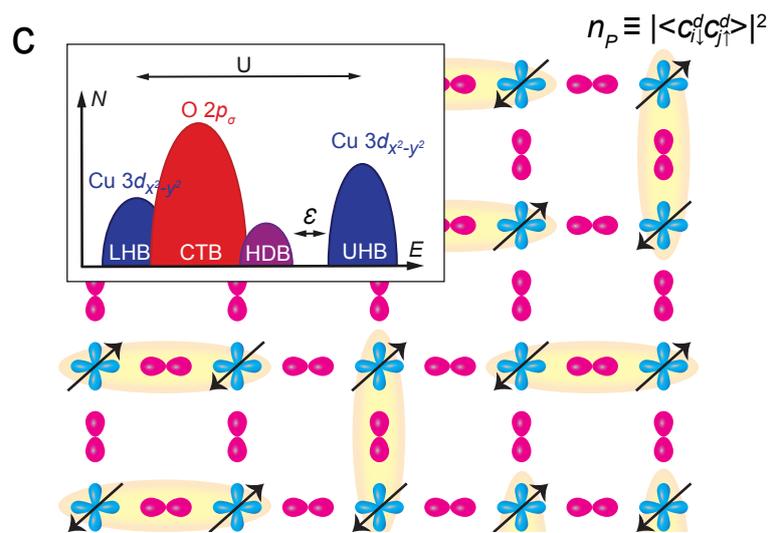

c

$$n_P \equiv |\langle c_i^d c_j^d \rangle|^2$$

d

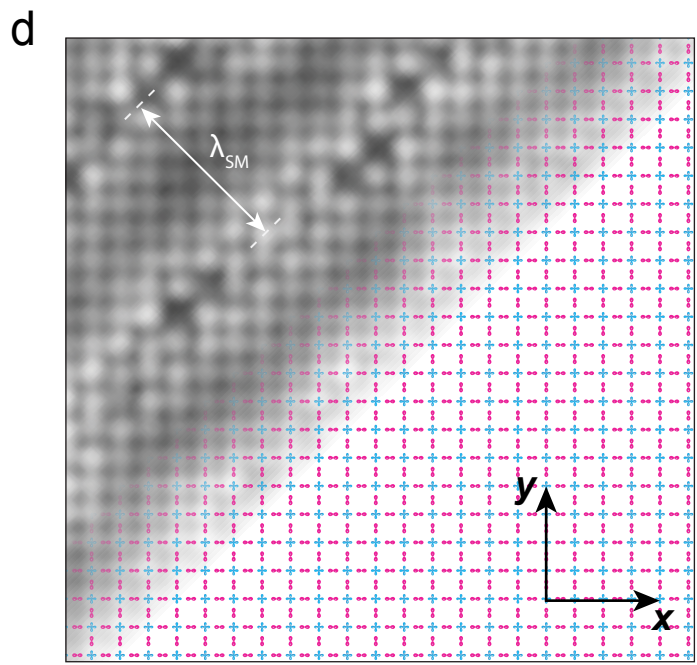

**Figure 1: Superexchange Magnetic Interactions in Transition-metal Oxides**

a. Schematic representation of CuO$_5$ pyramids whose bases comprise the CuO$_2$ plane. The degeneracy of Cu $d_{x^2-y^2}$ orbitals (blue) is lifted by the Hubbard energy U, and the O $p_\sigma$ orbitals (red) are separated from the upper Cu $d_{x^2-y^2}$ band by the charge transfer energy $\mathcal{E}$ (for holes).

b. Schematic of antiferromagnetic charge-transfer insulator state in undoped CuO$_2$. Inset shows a schematic density of electronic states N(E) in this phase, with the Coulomb energy U and the charge transfer energy $\mathcal{E}$ indicated. LHB, lower Hubbard band. UHB, upper Hubbard band. CTB, charge-transfer band.

c. Schematic of hole-doped CuO$_2$, a two-dimensional correlated metallic state with intense antiferromagnetic spin-spin interactions. When superconductive, the electron-pair condensate $\Psi \equiv \langle c_{i\downarrow}^d c_{j\uparrow}^d \rangle$ is indicated schematically in yellow, and the related electron-pair density is $n_P \equiv \left| \langle c_{i\downarrow}^d c_{j\uparrow}^d \rangle \right|^2$. Inset shows a schematic N(E) in this phase which, although reorganized by the delocalized carriers, still retains a charge-transfer energy scale $\mathcal{E}$. HDB, hole-doped band.

d. Schematic of CuO$_2$ partially overlaid by a Bi$_2$Sr$_2$CaCu$_2$O$_{8+x}$ topographic image $T(\boldsymbol{r})$ to exemplify how the crystal supermodulation modulates along the (1,1) axis, with one period $0 \leq \Phi \leq 2\pi$ requiring approximately $26\,\text{Å}$. The Cu to apical O distance $\delta$ is modulated at same wavevector but perpendicular to this plane.

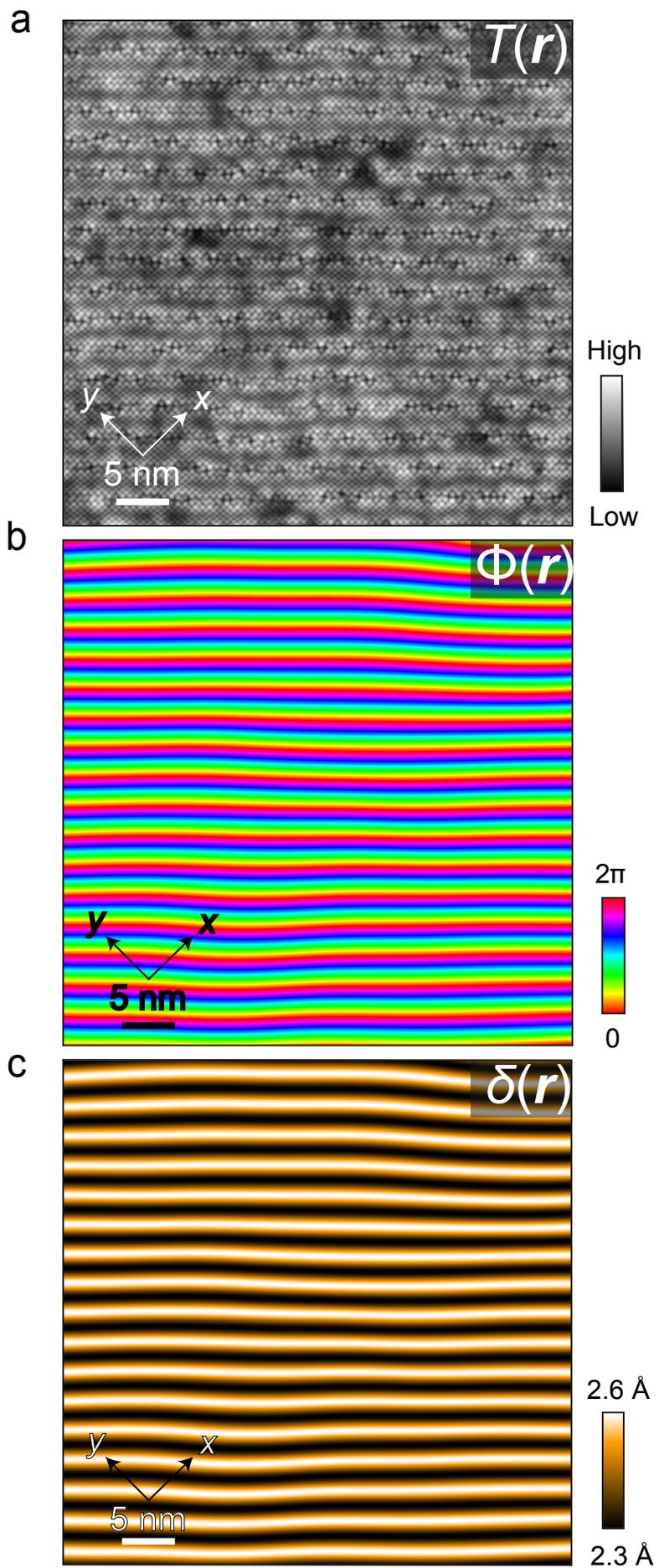

**Figure 2: Imaging Supermodulation Phase $\Phi(r)$ and Apical Oxygen Distance $\delta(r)$**

a. Exemplary $Bi_2Sr_2CaCu_2O_{8+x}$ topograph $T(r)$ at the BiO termination layer. The planar Cu-O axes are at 45-degrees to the supermodulation, as shown. The supermodulation runs from top to bottom with wavevector $Q_S \approx (0.15, 0.15) 2\pi/a_0$, obviously with relatively short correlation length.

b. From a, the supermodulation phase $\Phi(r)$ is derived (Ref. 10 Section II).

c. From b, the apical distance $\delta(r)$ is derived from X-ray refinement data for the $Bi_2Sr_2CaCu_2O_{8+x}$ crystal structure (Ref. 10 Section II).

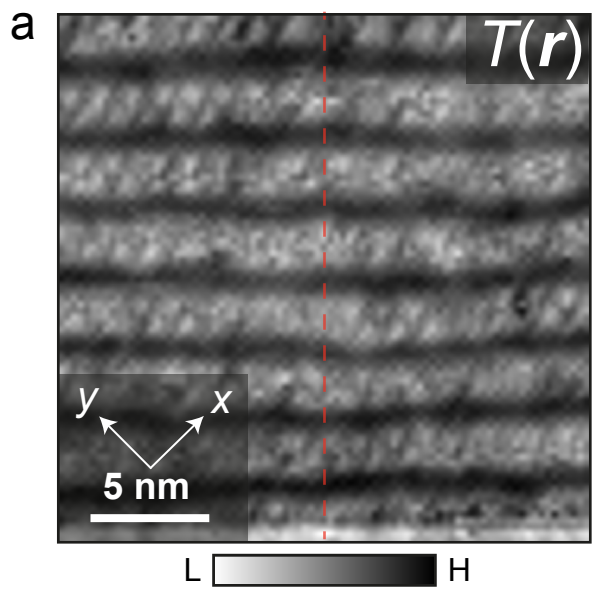

**a** $T(\boldsymbol{r})$

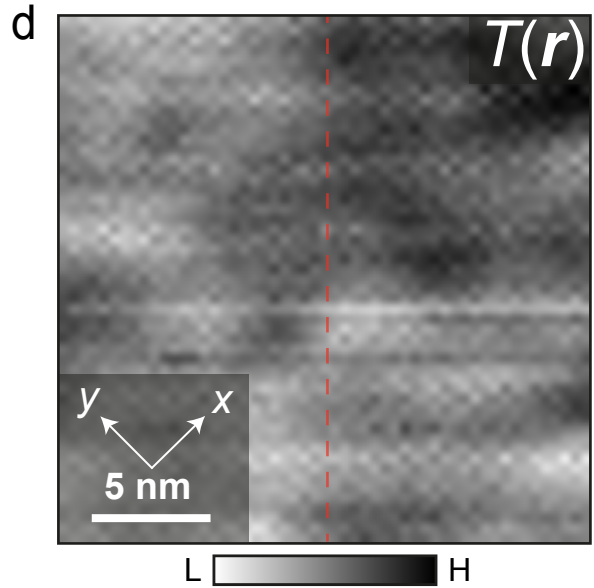

**d** $T(\boldsymbol{r})$

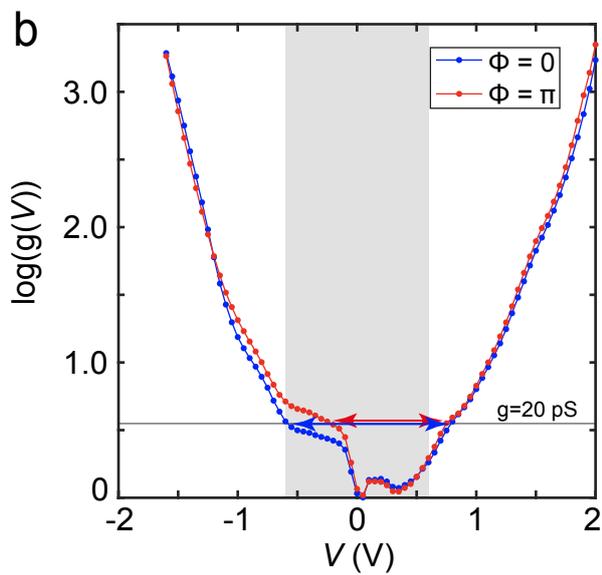

**b**

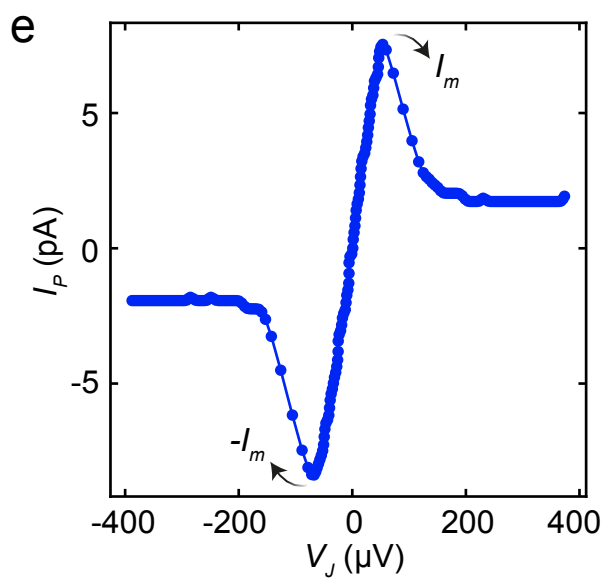

**e**

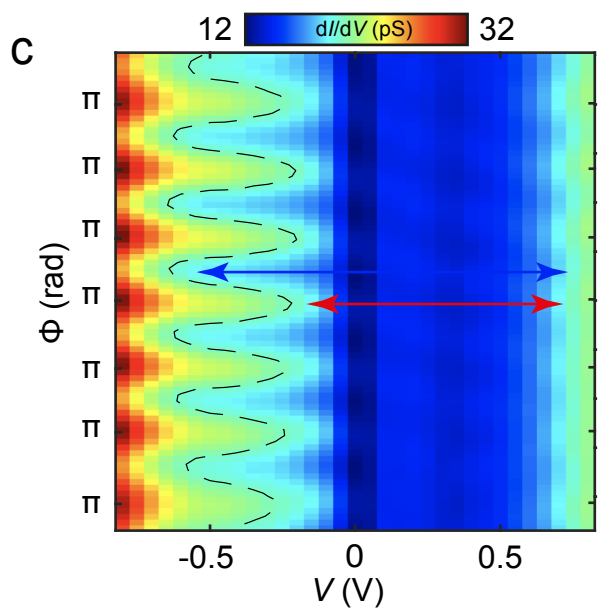

**c**

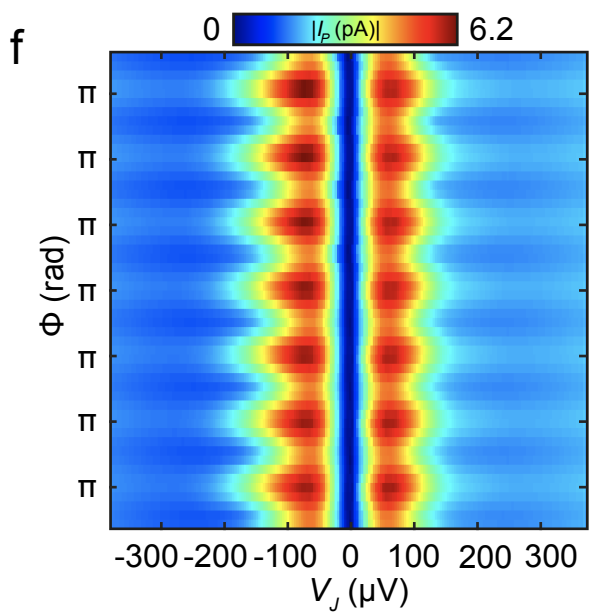

**f**

**Figure 3: Visualizing Charge-Transfer Energy $\mathcal{E}$ and Electron-Pair Density $n_P$**

a. Topographic image of BiO termination layer at T=4.2K, using a non-superconducting W-tip. Trajectory of dashed red line corresponds to the data in c.

b. $g(V)$ spectra of single-electron tunnelling measured at high-voltage and high tunnel junction resistance $R_N \approx 85\,\text{G}\Omega$ in the FOV of a averaged at supermodulation phases $\Phi = 0$ and $\Phi = \pi$. Use of logarithmic scale $\log(g(V))$ reveals exponential growth of density of states away from gap edges (*29*). The estimated value of $\mathcal{E}$ is derived as the minimum energy separation between the bands at constant $g = 20\,\text{pS}$, as shown by double-head arrows. The value of $\mathcal{E}$ is shown to change by $\approx 0.3$ eV from $\Phi = 0$ to $\Phi = \pi$. (Ref 10, Section III).

c. Measured $g(V)$ along the dashed line in a. The energy difference $\mathcal{E}$ between the lower and upper gap edge is very clearly modulating, with typical examples of $\mathcal{E}(\Phi = 0)$ and $\mathcal{E}(\Phi = \pi)$ indicated by arrows blue and red double-headed arrows, respectively.

d. Topographic image of BiO termination layer at T=2.1K, using a superconducting tip. Trajectory of dashed red line corresponds to the data in f.

e. Typical $I_P(V_J)$ spectrum of electron-pair tunnelling measured at low-voltage and $R_N \approx 21$ M$\Omega$ in the FOV of d.

f. Measured $|I_P(V_J)|$ along the dashed linecut in d. The maxima of the electron-pair current are very clearly modulating at the same wavevector as in c. Though not a direct measure of $n_p(r)$, this gives the most direct empirical indication that supermodulations are occurring in the pair density. The minima(maxima) in $|I_P(V_J)|$ occur at $\Phi = m2\pi$ ($\Phi = (2m + 1)\pi$) where m is an integer. We note that it is the maxima(minima) in the pseudogap energy as measured by single-particle tunnelling, that occur at the equivalent phases of the supermodulation[25], as might be expected from the relationship between pseudogap and condensed pair density in the cuprate phase diagram. For clarity, c. and f. have been Fourier filtered at the crystal supermodulation wavevector.

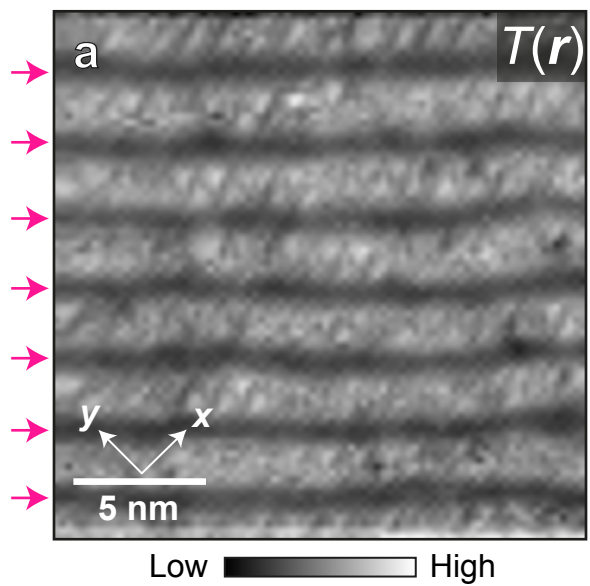

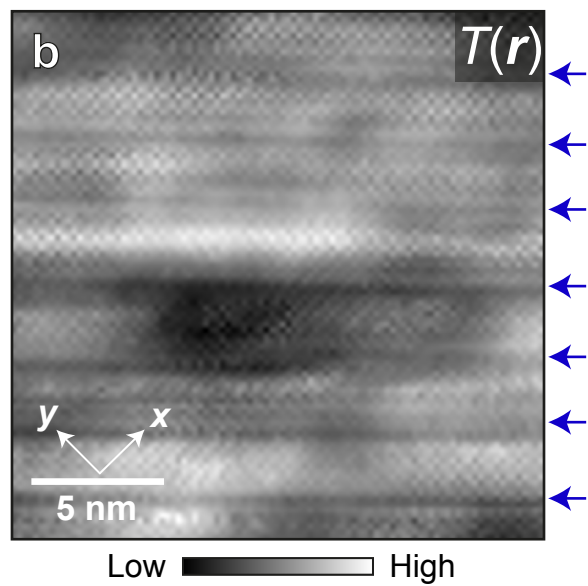

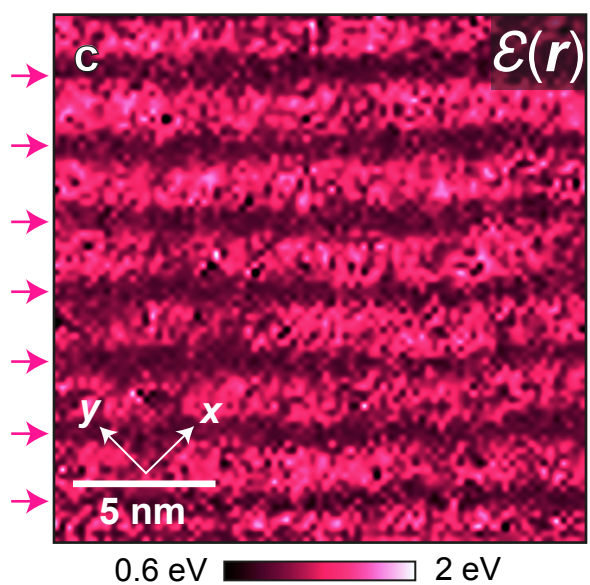

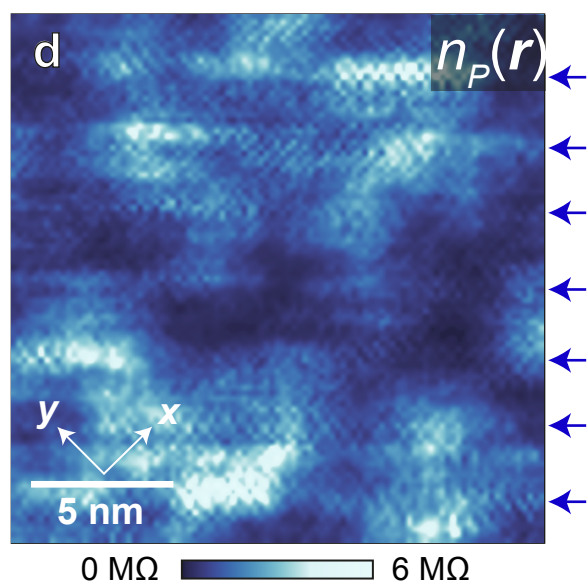

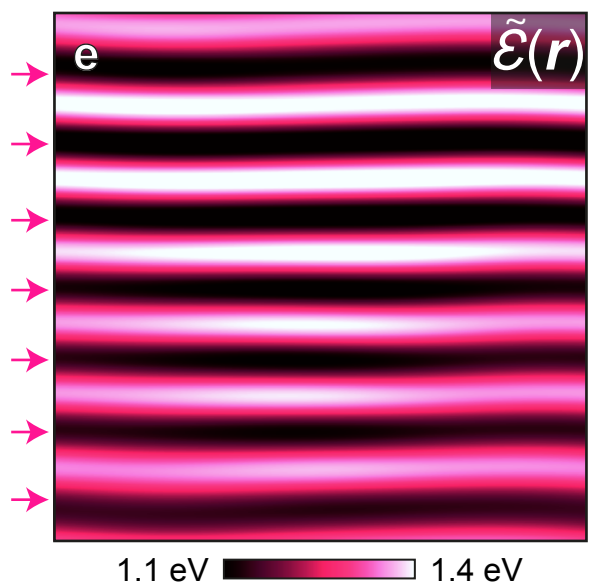

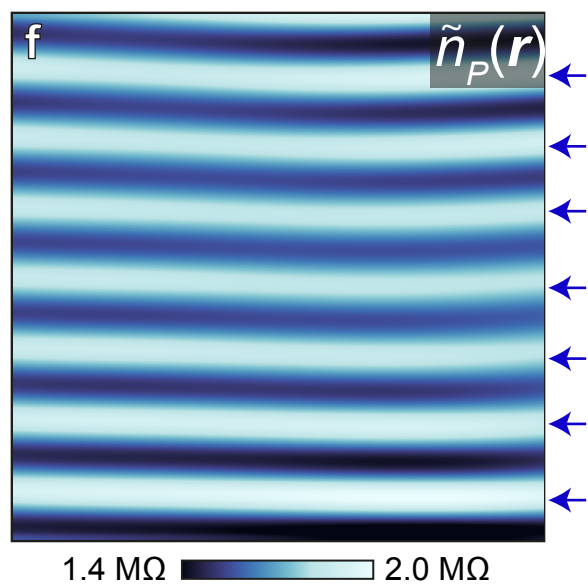

**Figure 4: Atomic-scale Visualization of $\mathcal{E}(\boldsymbol{r})$ and $\boldsymbol{n_P(r)}$ Versus $\boldsymbol{\delta(r)}$**

a. Topographic image $T(\boldsymbol{r})$ simultaneous with high-voltage $g(\boldsymbol{r}, V)$ measured at $R_N \approx 85\,\text{G}\Omega$ yielding c. The pink arrowheads are at supermodulation $\Phi = \pi$ as determined using the procedures described in Ref 10 Section II.

b. Topographic image $T(\boldsymbol{r})$ simultaneous with low-voltage $I_P(\boldsymbol{r}, V_J)$ and $R_N(\boldsymbol{r})$ maps yielding d. The blue arrowheads are at $\Phi = \pi$ as determined using the procedures described in Ref 10 Section II. The topographic image has atomic resolution allowing the BiO layer to be discerned clearly, although it is somewhat different than a, due to use of a Bi$_2$Sr$_2$CaCu$_2$O$_{8+x}$ nanoflake superconductive tip (7) (Ref 10 Section IV).

c. Measured $\mathcal{E}(\boldsymbol{r})$ in the FOV of a. The mean value is $\mathcal{E} = 1.195\,\text{eV}$, which is in very good agreement with $\mathcal{E}(\boldsymbol{r})$ for Bi$_2$Sr$_2$CaCu$_2$O$_{8+x}$ derived independently from other techniques (Ref. 10 Section III). The pink arrowheads are at $\Phi = \pi$ of the supermodulation.

d. Measured $n_P(\boldsymbol{r})$ in the FOV of b. (Ref 10 Section IV). The blue arrowheads are at $\Phi = \pi$.

e. Fourier filtered $\tilde{\mathcal{E}}(\boldsymbol{r})$ at supermodulation wavevectors $\pm \boldsymbol{Q_S}$ in the FOV of a,c. The pink arrowheads are at $\Phi = \pi$.

f. Fourier filtered $\tilde{n}_P(\boldsymbol{r})$ at supermodulation wavevectors $\pm \boldsymbol{Q_S}$ in the FOV of b,d. The blue arrowheads are at $\Phi = \pi$ .

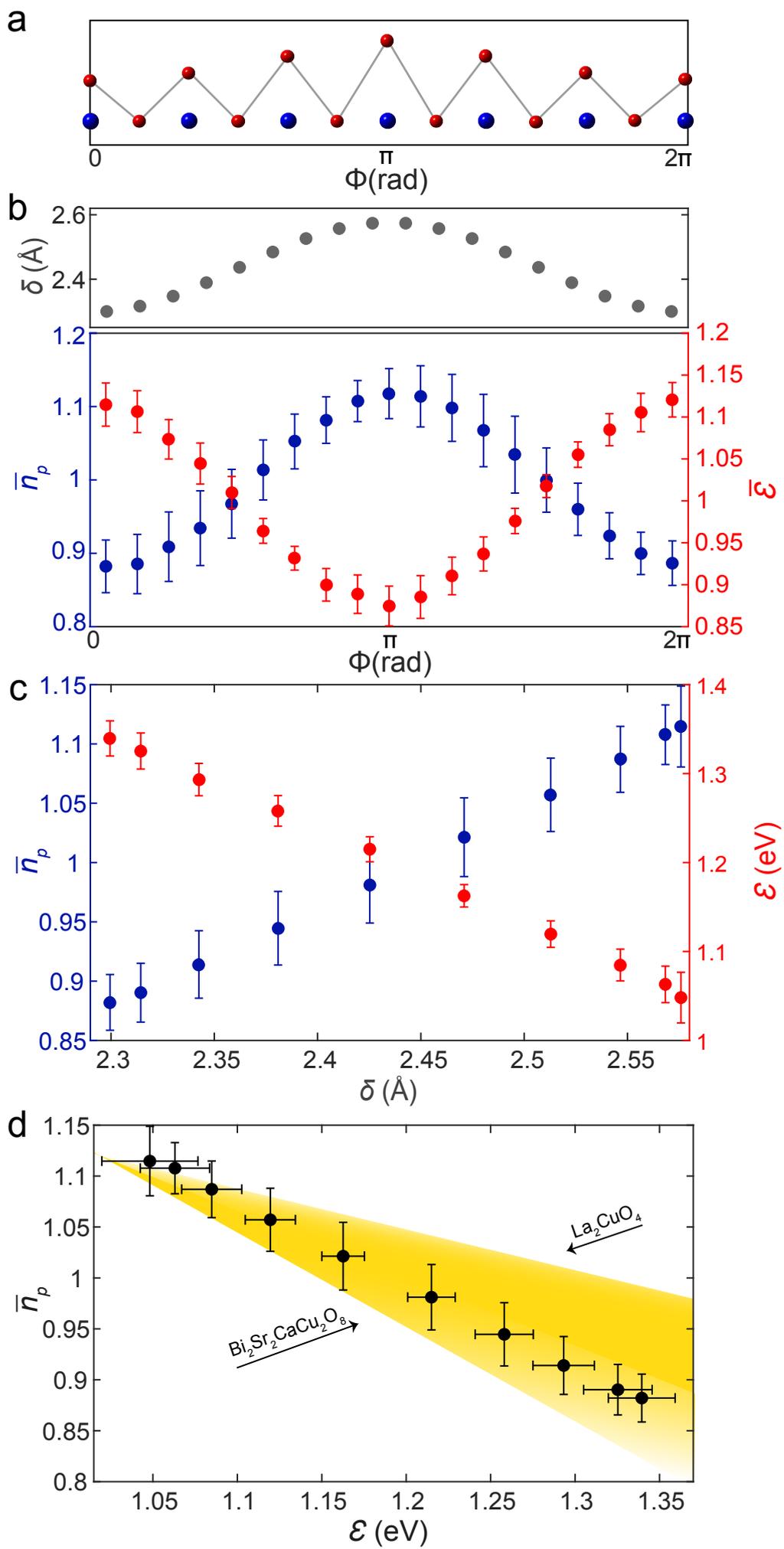

**Figure 5: Evolution of Cuprate Electron-pair Density $n_P$ with Charge-Transfer Gap $\mathcal{E}$**

a. Schematic of planar Cu to apical O distance modulations $\delta(\boldsymbol{r})$ in Bi$_2$Sr$_2$CaCu$_2$O$_{8+x}$ shown versus supermodulation phase $\Phi$.

b. Grey dots: $\delta(\Phi)$ showing the displacement of the apical oxygen atom within the CuO$_5$ pyramid versus supermodulation phase $\Phi$ (24).

   Red dots: measured $\bar{\mathcal{E}}(\Phi)$ showing the typical value for the Cu-O charge-transfer energy $\mathcal{E}$ for each value of the supermodulation phase $\Phi$ normalised to the mean value of $\mathcal{E}$. These data are from the same FOV as 4 a,c,e. Blue dots: measured $\bar{n}_P(\Phi)$ showing the measured value of electron-pair density versus supermodulation phase $\Phi$. These data are from a larger FOV comprising 13 supermodulation periods which contains the FOV from 4 b,d,f.

c. Measured dependence of Cu-O charge-transfer energy $\mathcal{E}$ and electron-pair density $n_P$ on the displacement $\delta$ of the apical O atoms from the planar Cu atoms.

d. Measured relationship of electron-pair density $\bar{n}_P$ to the Cu-O charge-transfer energy $\mathcal{E}$ in the CuO$_2$ plane of Bi$_2$Sr$_2$CaCu$_2$O$_{8+x}$. The yellow shaded region shows the range of predicted slopes for $d\bar{n}_P/d\mathcal{E} \equiv -\alpha$, as $0.3 \lesssim \alpha \lesssim 1.0 \text{ eV}^{-1}$ . These are derived from DMFT calculations for various materials with the limits reported for La$_2$CuO$_4$ and Bi$_2$Sr$_2$CaCu$_2$O$_{8+x}$ as indicated by black arrows. Error bars for b,c and d are obtained from the standard deviation of the phase-averaged values.

**Acknowledgements**: We acknowledge and thank S.D. Edkins, K. Fujita and J. Zaanen for helpful discussions and advice. We are grateful to D.-H. Lee for critical theoretical guidance on the interplay of superexchange with the electron-pair orders. M.H.H, X.L., Y.X.C., and J.C.S.D acknowledge support from the Moore Foundation's EPiQS Initiative through Grant GBMF9457. S.O'M and J.C.S.D. acknowledge support from Science Foundation of Ireland under Award SFI 17/RP/5445. W.C. and J.C.S.D. acknowledge support from the Royal Society under Award R64897. W.R. and J.C.S.D. acknowledge support from the European Research Council (ERC) under Award DLV-788932.

**Author Contributions:** J.C.S.D. designed research; S.O.M., W.R., Y.X.C., X.L., W.C., and M.H.H. performed research; H.E. and S.U. contributed new reagents/analytic tools; H.E. and S.U. synthesized and characterized the samples; M.H.H. carried out SJTM measurements and data processing; S.O.M., W.R., W.C., Y.X.C., and X.L. analyzed data; and S.O.M., W.R., W.C., and J.C.S.D. wrote the paper.

**Author Information** Correspondence and requests for materials should be addressed to jcseamusdavis@gmail.com.

Supplementary Material for

On the Electron Pairing Mechanism of
Copper-Oxide High Temperature Superconductivity

S.M. O'Mahony, Wangping Ren, Weijiong Chen, Yi Xue Chong, Xiaolong Liu,
H. Eisaki, S. Uchida, M.H. Hamidian, and J.C. Séamus Davis

## (I) Charge Transfer Energy, Superexchange and Superconductivity

Many types of studies of $\mathcal{E}$ and its consequent $J$ for cuprates have been carried out on bulk crystals. Optical reflectivity (*46,47*) revealed that the charge transfer energy $\mathcal{E}$ in the parent insulating state (Fig. 1b), for virtually all cuprates that become superconducting under hole-doping, is $1\ eV < \mathcal{E} < 2\ eV$. Raman scattering (*46,47,48*) identified the concomitant electron-pair excitations as a superexchange energy in the range $J{\sim}150\ meV$. Single-electron tunneling spectroscopy (*27*) reveals that the charge transfer energy for $Bi_2Sr_2CaCu_2O_{8+x}$ in the insulating parent state is $\mathcal{E} \gtrsim 1\ eV$, and a non-monotonic but diminishing trend of maximum $T_C$ measured in the superconductor with increasing $\mathcal{E}$ measured in the corresponding insulator, for samples of several distinct material types. Advanced angle resolved photoemission (*49*) reports transitions to unoccupied states consistent with a Cu on-site Coulomb interaction strength U of 2.7 eV and that the charge-transfer energy of optimally doped $Bi_2Sr_2CaCu_2O_{8+x}$ is then $\mathcal{E} \approx 1.1\ eV$. Resonant inelastic X-ray scattering (*50,51*) reveals directly, from the spectrum of spin wave excitations of many cuprate materials, that $140\ meV < J < 180\ meV$. Most recently (*52*) X-ray absorption spectra at both the Cu-L edge and O-K edge were used to determine $\mathcal{E} \approx 1.3\ eV$ in optimally superconducting $(Ca_xLa_{1-x})(Ba_{1.75-x}La_{0.25+x})Cu_3O_y$. Nevertheless, despite all these pioneering studies of $\mathcal{E}$ and $J$ for cuprates, simultaneous measurements of effects on the cuprate electron-pair condensate amplitude $\langle c_i^{\dagger} c_j^{\dagger} \rangle$ of either $\mathcal{E}$ or $J$ have never, to our knowledge, been reported.

## (II) Supermodulation Phase $\Phi(r)$ and Apical Displacement $\delta(r)$ Estimation

The bulk crystal supermodulation in $Bi_2Sr_2CaCu_2O_{8+x}$ perturbs the atoms from their idealized crystal positions due to the mismatch between the bond lengths of the rock-salt layer and the perovskite layer of $Bi_2Sr_2CaCu_2O_{8+x}$. The modulation has a periodicity of $\sim 26\ Å$ along the *a*-axis of the crystal (*53*) but lacks long range coherence. The distance between all atoms in each $Bi_2Sr_2CaCu_2O_{8+x}$ unit cell modulates periodically, with the change in distance between the planar Cu atom and the apical O atom being largest in amplitude (Fig. S1) varying between $2.25\ Å$ and $2.54\ Å$ over one supermodulation period (Fig 5a). X-ray scattering data also reveals that the Bi-O distance is virtually constant over the supermodulation period (Fig. S1), enabling direct connection between the supermodulation signal in the topographic image $T(r)$ and the modulation of the planar Cu to apical O separation $\delta(r)$ (Fig. S1). Fig. S1 also shows the displacement of the BiO termination layer from the symmetry plane of the crystal, which is equivalent to a topograph $T(\boldsymbol{r})$. This allows one to directly relate the supermodulation phase $\Phi$ obtained from $T(\boldsymbol{r})$ to the X-ray refinement data (53).

Effects of the supermodulation can be observed in topographic images $T(\boldsymbol{r})$ measured at the BiO crystal termination layer of $Bi_2Sr_2CaCu_2O_{8+x}$. Its signal as a component of the topographic image can be expressed as

$$T'(\boldsymbol{r}) = A_S(\boldsymbol{r}) \cos \Phi(\boldsymbol{r}), \tag{S1}$$

$$\Phi(\boldsymbol{r}) = \boldsymbol{Q}_S \cdot \boldsymbol{r} + \theta(\boldsymbol{r}), \tag{S2}$$

where $\boldsymbol{Q}_S$ is the supermodulation vector and $\theta(\boldsymbol{r})$ is a spatially dependent phase disorder. To measure $\Phi(\boldsymbol{r})$ we process $T(\boldsymbol{r})$ retaining only the wavevector components near $\boldsymbol{q} = \pm\boldsymbol{Q}_S$. In practice, this is achieved by first Fourier filtering $T(\boldsymbol{r})$ as follows:

$$T'(\boldsymbol{r}) = \frac{1}{\sigma\sqrt{2\pi}} \int T(\boldsymbol{q}) \left( e^{\frac{|\boldsymbol{q}-\boldsymbol{Q}_S|^2}{2\sigma^2}} + e^{\frac{|\boldsymbol{q}+\boldsymbol{Q}_S|^2}{2\sigma^2}} \right) \exp(-i\boldsymbol{q} \cdot \boldsymbol{r}) \, \mathrm{d}\boldsymbol{q} \tag{S3}$$

where $T(\boldsymbol{q})$ is the Fourier transform of $T(\boldsymbol{r})$. We use $\sigma \approx \frac{2\pi}{20}$ nm$^{-1}$ when filtering $\mathcal{E}(\boldsymbol{r})$ and $\sigma \approx \frac{2\pi}{40}$ nm$^{-1}$ when filtering $n_P(\boldsymbol{r})$, to take account of the difference in tip sizes. Next, to visualize the phase disorder $\theta(\boldsymbol{r})$, we employ a two-dimensional lock-in method in which references $\alpha(\boldsymbol{r}) = \sin(\boldsymbol{Q}_S \cdot \boldsymbol{r})$ and $\beta(\boldsymbol{r}) = \cos(\boldsymbol{Q}_S \cdot \boldsymbol{r})$ are multiplied by $T'(\boldsymbol{r})$:

$$X(\boldsymbol{r}) \equiv T'(\boldsymbol{r})\alpha(\boldsymbol{r}), \tag{S4}$$

$$Y(\boldsymbol{r}) \equiv T'(\boldsymbol{r})\beta(\boldsymbol{r}). \tag{S5}$$

Low pass filtering these product images to remove the AC term, yields

$$\theta(\boldsymbol{r}) = \arctan\left(\frac{Y(\boldsymbol{r})}{X(\boldsymbol{r})}\right). \tag{S6}$$

Adding back the pure modulating $\boldsymbol{Q}_S \cdot \boldsymbol{r}$ term and defining the result modulo $2\pi$, generates a map of the total supermodulation phase $\Phi(\boldsymbol{r}) = [\boldsymbol{Q}_S \cdot \boldsymbol{r} + \theta(\boldsymbol{r})] \bmod 2\pi$.

The relationship between the apical Cu-O displacement $\delta$ and phase $\Phi$ can then be obtained using X-ray refinement techniques (53) that report apical displacement varies between $\delta \sim 2.25$ Å at $\Phi = 0$ and $\delta \sim 2.54$ Å at $\Phi = \pi$ (Fig. S1) The first harmonic is reported as:

$$\delta(\Phi) = 2.44 - 0.14 \cos(\Phi) \text{ Å} \tag{S7}$$

This enables to make direct connection between the measured supermodulation phase $\Phi(\boldsymbol{r})$ obtained from the topographic image $T(\boldsymbol{r})$ and the Cu-O separation $\delta(\boldsymbol{r})$ within each crystal unit cell. Fig. S2 shows a typical example of the equivalent $T(\boldsymbol{r}):\Phi(\boldsymbol{r}):\delta(\boldsymbol{r})$ arising from this procedure when using a $Bi_2Sr_2CaCu_2O_{8+x}$ nanoflake tip for SJTM measurements of $n_P(\boldsymbol{r})$.

## (III) Charge Transfer Energy Visualization $\mathcal{E}(r)$ by SISTM

To extract the spatially resolved charge transfer energy $\mathcal{E}(\boldsymbol{r})$, high voltage (26,54,55,) differential tunnel conductance $g(\boldsymbol{r}, V)$ imaging of $Bi_2Sr_2CaCu_2O_{8+x}$ is used. Here, we visualize $g(\boldsymbol{r}, V)$ in the range $-1.6 \text{ V} \leq V \leq 2 \text{ V}$, and at the extremely high junction resistance $R_N \approx 85$ G$\Omega$ ($V_S = 600$ mV: $I_S = 7$pA) necessary to suppress all tip-induced electric field effects. Fig. S3g then shows a typical example of a $g(\boldsymbol{r}, V)$ series measured at T=4.2K along a line transverse to the crystal supermodulation. The edges of the filled lower band and the empty upper band can be identified from the appearance of extremely rapid increase in the density of states. Here, as Fig S3g, we produce Fig. 3c from the main text showing the measured $g(\boldsymbol{r}, V)$ in which these $g(\boldsymbol{r}, V)$ variations at the periodicity as the supermodulation are clearly evident.

One must consider the possibility of a "setup effect" on these data. Most simply, the electron tunneling current from the STM tip to the sample is

$$I(\boldsymbol{r}, V) = C e^{-\frac{T(\boldsymbol{r})}{T_0}} \int_0^{E=eV} [f(E, T) N(\boldsymbol{r}, E)] \big[ (1 - f(E, T)) N_{\text{Tip}}(E) \big] dE \qquad (S8)$$

where $T(\boldsymbol{r})$ is the 'topograph', $V$ the tip-sample bias voltage, $N(\boldsymbol{r}, E)$ the sample's local-density-of-electronic-states, $N_{\text{TIP}}(E)$ the tip density-of-electronic-states $f(E,T)$ is the Fermi function. Here $C e^{-\frac{T(\boldsymbol{r})}{T_0}}$ contains all the effects of tip elevation, tip-sample work-functions and tunneling matrix element. Thus, at very low temperatures and with $N_{\text{TIP}}(E)$ and $C(\boldsymbol{r})$ both constants

$$I_s = C e^{-\frac{T(\boldsymbol{r})}{T_0}} \int_0^{eV_s} N(\boldsymbol{r}, E) dE \Rightarrow C e^{-\frac{T(\boldsymbol{r})}{T_0}} = I_s / \int_0^{eV_s} N(\boldsymbol{r}, E) dE \qquad (S9)$$

where $V_S$ and $I_S$ are the (constant but arbitrary) 'set-up' bias voltage and current respectively. In practice, these two parameters fix the tip elevation $T(\boldsymbol{r})$ for each tip-sample junction. Equivalently, standard constant-current topographic imaging adjusts $T(\boldsymbol{r})$ as the tip scans over the sample surface to maintain a set point current, $I_S$, at a constant applied tip-sample bias $V_S$. In this case, the topographic image $T(\boldsymbol{r}, V_S)$ is

$$T(\boldsymbol{r}, V_S) = T_0 \ln \left[ \int_0^{E=eV_S} N(\boldsymbol{r}, E) dE \right] \qquad (S10)$$

up to a constant. The energy derivative of Eq. S8 at zero temperature yields the tip-sample differential conductance $dI/dV(\boldsymbol{r}, E = eV) \equiv g(\boldsymbol{r}, E)$ which, upon substitution of $C e^{-\frac{T(\boldsymbol{r})}{T_0}}$ from Eq. S9 yields

$$g(\boldsymbol{r}, V) = e I_s N(\boldsymbol{r}, E) / \int_0^{eV_s} N(\boldsymbol{r}, E) dE \qquad (S11)$$

This is the infamous "setup effect" in which $g(\boldsymbol{r}, V)$ is not proportional to $N(\boldsymbol{r}, E)$. Any setup effect of the $\text{Bi}_2\text{Sr}_2\text{CaCu}_2\text{O}_{8+x}$ crystal supermodulation would then be to modulate $\int_0^{eV_s} N(\boldsymbol{r}, E) dE$ periodically at wavevector $\boldsymbol{Q}_S$. One can simulate the effects which this would produce by multiplying a typical $g(\boldsymbol{r}, V)$ by a periodic function $A\text{Cos}(\boldsymbol{Q}_S. \boldsymbol{r})$ and this we have done to generate simulated data in Fig. S3j. Hence, if a setup effect were the predominant phenomenon, the measured $g(\boldsymbol{r}, V)$ data in Fig. S3g would correspond to the image of setup-effect simulated $g(\boldsymbol{r}, V)$ data in Fig S3j. Inspection shows them be quite different, most obviously in that there is virtually no modulation in measured $g(\boldsymbol{r}, V)$ at large positive bias voltage, as would be required to occur if the setup-effect prevails. This analysis rules out the setup effect (Eq. S11) as the predominant cause of modulations at $\boldsymbol{Q}_S$ in all $g(\boldsymbol{r}, V)$ data studied herein.

Hence, variations in the energy separation $\mathcal{E}$ between the bands can be estimated by measuring the energy range between these edges at a constant differential conductance $G$ as shown, for example, in Fig. 3c of the main text. In our studies we choose to use $G \approx 20$ pS for several practical reasons explained below, but emphasize that the demonstration that $\mathcal{E}(\boldsymbol{r})$ varies periodically with the crystal supermodulation is not contingent on any specific value for $G$. For example, in Fig. S4 shows $\mathcal{E}(G, \boldsymbol{r})$ maps extracted at selected values of $G$ ranging from 20 pS to 80 pS. One can see that the spatial structure of $\mathcal{E}(G, \boldsymbol{r})$ is extremely similar for 20 pS $< G <$ 80 pS. Moreover, the power spectral density Fourier transform $\mathcal{E}(G, \boldsymbol{q})$ of every one exhibits strong peaks at wavevectors $\pm \boldsymbol{Q}_S$. This means that the relationship between $\mathcal{E}$ and $\Phi$ will be qualitatively the same for any $G$ considered here, but with somewhat different amplitudes in $\mathcal{E}$ variations. One additional complication for the $\mathcal{E}(G, \boldsymbol{r})$ measured in the range 40 pS $< G <$ 80 pS and that at $G = 20$ pS is the presence of dark regions at each oxygen dopant ion where strong local maxima occur in $g(\boldsymbol{r}, V)$ at conductance higher than 20 pS, preventing accurate measurement of $\mathcal{E}(\Phi)$. This is another reason we focus on analysis of the $G = 20$ pS charge transfer energy maps so as to avoid the dopant-atom complication. Incidentally, we demonstrate directly in Figs S3b,c; S3e,f; S3h,i; S3k,l that the crystal supermodulation effects on $\mathcal{E}$ occur in regions where dopant ions are absent, ruling them out as a possible contribution to the phenomena under study.

Finally, Fig. S5 (a-c) show $\mathcal{E}(\boldsymbol{r})$ obtained by this $G = 20$ pS procedure in three disjoint fields of view. Each has an area of $19.5 \times 19.5$ nm$^2$ and their $\mathcal{E}(\boldsymbol{r})$ have extremely similar spatial structure, with spatially-averaged values $\langle \mathcal{E} \rangle \approx$ 1.195 eV, 1.205 eV and 1.215 eV respectively. These average charge-transfer energies are in excellent quantitative agreement with those derived from a variety of experimental techniques, including SISTM (27), ARPES (56), and optical spectroscopy (57) of Bi$_2$Sr$_2$CaCu$_2$O$_{8+x}$, meaning that the choice of $G = 20$ pS at which to measure $\mathcal{E}$ is both practical and quantitatively valid. To further underscore the statistical validity of $\mathcal{E}$, Fig. S5 d,e,f show the phase-averaged, Fourier-filtered $\mathcal{E}(\Phi)$ in the same three fields of view. It is readily apparent that the peak-to-peak amplitude is the same for all three fields of view, and $\mathcal{E}(\Phi)$ is minimized at $\Phi = \pi$ in all three fields of view. Ultimately, it is this band-separation at $G = 20$ pS analysis technique which yields the $\mathcal{E}(\Phi)$ that we study throughout the main text.

## (IV) Electron-pair Density Visualization $n_P(r)$ by SJTM

The macroscopic wavefunctions of the tip and sample forming a Josephson junction can be written as

$$\psi_T = \langle c_\uparrow c_\downarrow \rangle_T \mathrm{e}^{-\mathrm{i}\varphi_T} = \sqrt{n_T}\mathrm{e}^{-\mathrm{i}\varphi_T} \tag{S12}$$
$$\psi_S = \langle c_\uparrow c_\downarrow \rangle_P \mathrm{e}^{-\mathrm{i}\varphi_S} = \sqrt{n_P}\mathrm{e}^{-\mathrm{i}\varphi_S} \tag{S13}$$

where $n_T$ ($n_P$) and $\varphi_T$ ($\varphi_S$) are the electron-pair densities and the phases of the tip (sample), respectively. The product of the Josephson critical current $I_J$ with the junction's normal-state resistance $R_N$ is (58,59):

$$I_J R_N \propto \langle c_\uparrow c_\downarrow \rangle_P \langle c_\uparrow c_\downarrow \rangle_T \propto \sqrt{n_P}\sqrt{n_T}. \tag{S14}$$

Assuming $n_T$ to be constant, the sample electron-pair density $n_P$ can be measured as $n_P \propto \left(I_J R_N\right)^2$. In a phase-diffusive Josephson junction, the current – voltage $I_P(V_J)$ characteristic is described by the following equation with the maximum Josephson current ($I_m = \frac{I_J^2 Z}{4V_c}$) appearing at non-zero junction voltages $V_c$:

$$I_P(V_J) = \frac{1}{2} I_J^2 Z \, V_J / \left(V_J^2 + V_c^2\right) \qquad (S15)$$

Here, $Z$ is the impedance relevant to re-trapping of the diffusing phase. The first derivative of $I_P(V_J)$ at zero bias is

$$g_0 \equiv \left. \frac{dI_P}{dV_J} \right|_{V_J=0} = \frac{I_J^2 Z}{2V_c^2} \propto I_m \qquad (S16)$$

In scanned Josephson tunneling microscopy (SJTM) using superconductive STM tips (eg. $Bi_2Sr_2CaCu_2O_{8+x}$ nanoflake tip as used in this study), the sample electron-pair density can therefore be visualized (7,8,9) by measuring either $I_m(\boldsymbol{r})$ or $g_0(\boldsymbol{r})$ and also separately measuring $R_N^2(\boldsymbol{r})$ in the identical field of view yielding the electron-pair density as:

$$n_P(\boldsymbol{r}) \propto I_m R_N^2(\boldsymbol{r}) \propto g_0 R_N^2(\boldsymbol{r}). \qquad (S17)$$

For these studies, we use a $Bi_2Sr_2CaCu_2O_{8+x}$ nanoflake tip (7,60). A single crystal sample of $Bi_2Sr_2CaCu_2O_{8+x}$ at the hole density $p = 0.17 \pm 0.01$ and superconducting transition temperature $T_c = 91$ K is cleaved in cryogenic ultra-high vacuum. The nanoflake tip is then created by picking up a nanometer scale $Bi_2Sr_2CaCu_2O_{8+x}$ flake from the sample surface. A typical resulting topography is shown in Fig. S2 in which both the locations of individual Bi atoms at the BiO termination layer and the crystal supermodulation are all well resolved (7,60).

To extract the phase-diffusive Josephson differential conductance at zero bias $g_0$ from $g(V_J, \boldsymbol{r})$ we fit this Josephson junction differential conductance spectrum to the 1st derivative of Eq. S15 with free fitting parameters being $\frac{1}{2} I_J^2 Z$, $V_c$ and an additional constant $C$ added to take account of any small conductance offsets that may occur systematically, and from which fits the $g_0(\boldsymbol{r})$ and $I_m(\boldsymbol{r})$ are derived.

Fig. S10 shows a linecut of $I_m$ along the $q = \boldsymbol{Q}_{SM}$ direction, normalized to the background value $I_m(\boldsymbol{q} = \boldsymbol{0})$ for three different experiments in three different fields of view along with the equivalent linecut for $g_0$ used to generate Fig. 5 of the main text. The consistency in the amplitude of supermodulation in $I_m$ and $g_0$ across the various experiments demonstrates the repeatability of the superfluid density modulation measurements.

## (V) Determination of Equivalent Normal State Junction Resistance $R_N$

The Josephson tunnel junction resistance $R_N(\boldsymbol{r})$ is difficult to determine in a single $g(\boldsymbol{r}, V)$ map because $I_P(V_J)$ is measured in the 50 μV range (Fig. 3b), whereas $R_N(\boldsymbol{r})$ is only quantifiable using differential conductance measurements at the 100 mV range (at $E \gg \Delta_{SIS}$ where for cuprates $\Delta_{SIS} > 50$meV). Hence we use a two-step procedure in which we measure a low voltage range $g_1(\boldsymbol{r}, V)$ and a high voltage range $g_2(\boldsymbol{r}, V)$ in exactly the same FOV (Fig. S6). The $g_1(\boldsymbol{r}, V)$ ranges from $-15$ mV $< V < 15$ mV with a setpoint $I_{s1} = 100$ pA, $V_{s1} = 15$mV, while $g_2(\boldsymbol{r}, V)$ is acquired from $-350$ mV $< V < 350$ mV in the almost identical FOV at setpoint $I_{s2} = 350$ pA, $V_{s2} = 350$mV so that $V_{s2} \gg \Delta_{SIS}$.

Thus, $g_1(\boldsymbol{r}, V)$ and $g_2(\boldsymbol{r}, V)$ are obtained in different measurements but nominally the same field of view (FOV). These images, along with their associated topographs $T(\boldsymbol{r})$, are then registered to the exact same FOV by a series of transformations. Each experimental image $T(\boldsymbol{r})$ can be registered to the expected periodic lattice using the Lawlor-Fujita (LF) procedure (*61*), which utilizes a 2D lock-in method to solve for the displacement field $\boldsymbol{r} - \tilde{\boldsymbol{r}}$ between the experimentally measured $T(\boldsymbol{r})$ and a perfectly lattice periodic map $T^0(\boldsymbol{r})$. After this procedure, the lattice is almost exactly periodic, but can sometimes appear sheared from expected $C4$ symmetry. To correct this, we apply a shear transformation to the experimental data. Once the experimental images have been corrected, they are then atomically registered to the perfectly identical FOV by rigidly shifting the maps relative to one another. After this procedure, the $g_1(\boldsymbol{r}, V)$ and $g_2(\boldsymbol{r}, V)$ data are now in a precisely identical FOV.

The $g_1(\boldsymbol{r}, V_J)$ is then fit with a parabolic curve $g_1 = a_1 V_J^2 + b_1$ over the voltage range $[-V_{s1}, V_{s1}]$ at every $\boldsymbol{r}$. A small energy range within which Josephson electron-pair tunneling dominates surrounding $V_J = 0$ is omitted. Next $g_2(\boldsymbol{r}, V_J)$ is smoothed and fit with a parabolic curve $g_2 = a_2 V_J^2 + b_2$ over a voltage range $[-V_0, V_0]$, where $V_{s1} < V_0 < \Delta_{SIS}$. Finally, we determine the scaling factors between $g_2(\boldsymbol{r})$ and $g_1(\boldsymbol{r})$ via $g_1(\boldsymbol{r}) = g_2'(\boldsymbol{r}) = \alpha(\boldsymbol{r})g_2(\boldsymbol{r}) + \beta(\boldsymbol{r})$, where $\alpha = \frac{a_1}{a_2}, \beta = b_1 - b_2 \frac{a_1}{a_2}$.

$$g_2'(\boldsymbol{r}, V_J) = \alpha(\boldsymbol{r})g_2(\boldsymbol{r}, V_J) + \beta(\boldsymbol{r}). \tag{S18}$$

The rescaled $g_2'(V_J)$ and the original $g_1(V_J)$ when plotted on the same graph are now virtually indistinguishable (Fig. S6c), and the high-voltage conductance $g_2'(\boldsymbol{r}, V_J)$ associated with the junction resistance $R_N$ for the conditions of Josephson tunneling be read off for each $\boldsymbol{r}$. The rescaled $g_2'(\boldsymbol{r}, V)$ and the original $g_1(\boldsymbol{r}, V)$ images are shown in Fig. S7 demonstrating the fidelity of the rescaling process. Finally, we can establish the normal state junction resistance corresponding to the low-voltage $g_1(\boldsymbol{r}, V_J)$ map:

$$R_N(\boldsymbol{r}) \equiv \frac{1}{g_2'(r, V_{s2})}. \tag{S19}$$

A typical example of such data $R_N(\boldsymbol{r})$ which is used to establish $n_P(\boldsymbol{r}) \propto g_0 R_N^2(\boldsymbol{r})$ as in Figs 4,5 is shown in Fig. S8.

**(VI) Determination of the Relationships $\tilde{\mathcal{E}}(\Phi)$ and $\tilde{n}_P(\Phi)$**

To measure the dependence of charge transfer energy modulations $\mathcal{E}(\Phi)$ and electron $-$ pair density modulations $n_P(\Phi)$ on the supermodulation phase $\Phi$, the first step is to Fourier filter the quantity of interest $A(\mathbf{r})$, either $\mathcal{E}(\mathbf{r})$ or $n_P(\mathbf{r})$, retaining only contributions from the wavevectors close to $\mathbf{q} = \pm \mathbf{Q}_S$ and adding the constant background $\langle A \rangle$:

$$\tilde{A}(\mathbf{r}) = \langle A \rangle + \frac{1}{\sigma\sqrt{2\pi}} \int A(\mathbf{q}) e^{-i q \cdot r} \left( e^{\frac{|q-Q_S|^2}{2\sigma^2}} + e^{\frac{|q+Q_S|^2}{2\sigma^2}} \right) d\mathbf{q}. \qquad \text{(S20)}$$

Next, we calculate the supermodulation phase $\Phi(\mathbf{r})$ from the topograph $T(\mathbf{r})$ in the same FOV as $A(\mathbf{r})$. We then bin each pixel in the FOV according to its local phase $\Phi(\mathbf{r})$. Since each pixel is associated with a pair $\tilde{A}(\mathbf{r}) : \Phi(\mathbf{r})$, this yields a two-dimensional histogram showing the coincidence of these values. This is shown for $\tilde{\mathcal{E}}(\mathbf{r})$ and $\tilde{n}_P(\mathbf{r})$ in Fig. S9a and S9c, which are the core empirical data for this project. The ultimate step is to average $\tilde{A}(\mathbf{r})$ for every $\mathbf{r}$ with a given value of $\Phi$ yielding the plot $\tilde{A}(\Phi)$ as shown for $\tilde{\mathcal{E}}(\mathbf{r})$ and $\tilde{n}_P(\mathbf{r})$ in Fig. S9b, d. This procedure produces Fig 5b, c and d of the main paper. The error bars are obtained as the standard deviation of the quantity $\tilde{A}(\mathbf{r})$ within each phase bin.

For $Bi_2Sr_2CaCu_2O_{8+x}$ a similar procedure has been used to plot the energy gap (colloquially in cuprate studies the "pseudogap") for single-particle excitations $\Delta(\Phi)$ as a function of phase (25). The results are that $\Delta$ is also modulated at $\mathbf{Q}_S$ and that its value is minimized at $\Phi \approx \pi$. Therefore, the data in Fig. 5 imply $\Delta$ and $n_p$ are out of phase by $\pi$ for $Bi_2Sr_2CaCu_2O_{8+x}$. This is as would be expected in a strongly correlated unconventional superconductor described by the Hubbard model. In that case, it is well known that the pseudogap to single-particle excitations evolves oppositely (38,39) to the amplitude of the amplitude of electron-pair condensate $\Psi$.

## (VII) Estimation of $\alpha$ from Three-band $CuO_2$ Hubbard Model Calculations

Theoretical calculations of $\Psi$ and $\mathcal{E}$ have been carried out by numerical solution of the Emery three-band model within the framework of cluster dynamical mean field theory (CDMFT). One can extract predicted values of $\alpha = dn_P/d\mathcal{E}$ for comparison with the experimental results in Fig. 5d. Here the measured $g_0 R_N^2$ is proportional to $n_P$, but the constants of proportionality are unknown so one we cannot (yet) compare this quantity to $n_P = |\Psi|^2$ calculated using CDMFT in absolute units. Thus, one needs to normalize both the experimental results and the theoretical predictions in a consistent manner.

To do so, we first normalize measured $g_0 R_N^2$ to its mean value in the experimental FOV:

$$\bar{n}_P = \frac{g_0 R_N^2}{\langle g_0 R_N^2 \rangle} \qquad \text{(S21)}$$

as shown in Fig. 5. The CDMFT calculated $|\Psi|$ in Ref. (*12*) is normalized to the reference value of $|\Psi|$ for the so-called covalent case $|\Psi|_{cov} = 0.0774$, whose input material parameters are chosen to be representative of $Bi_2Sr_2CaCu_2O_{8+x}$.

$$\left|\overline{\langle c_\uparrow c_\downarrow \rangle}\right|^{BSCCO} = \frac{|\Psi|}{|\Psi|_{cov}} \tag{S22}$$

Normalization of results for $La_2CuO_4$ in Ref. (*16*) is achieved similarly as

$$\left|\overline{\langle c_\uparrow c_\downarrow \rangle}\right|^{LSCO} = \frac{|\Psi|}{|\Psi|_{LSCO}}, \tag{S23}$$

where $|\Psi|_{LSCO} = 0.0154$. Since $Bi_2Sr_2CaCu_2O_{8+x}$ is among the cuprates with the smallest value of $\mathcal{E}$ and $La_2CuO_4$ is among those with the largest, we define a range of $\alpha$ for the cuprates ranging between the values of Ref. (*12*) and Ref. (*16*) to arrive at the approximate inequality $0.3 \lesssim \alpha \lesssim 1.0 \text{ eV}^{-1}$. This is represented by the yellow shaded wedge in Fig. 5d. The measured value of $\alpha = -0.81 \pm 0.17 \text{ eV}^{-1}$ falls in this range and is very close to the DMFT prediction for $Bi_2Sr_2CaCu_2O_{8+x}$.

# Supplementary References

# Supplementary Figures

**a**

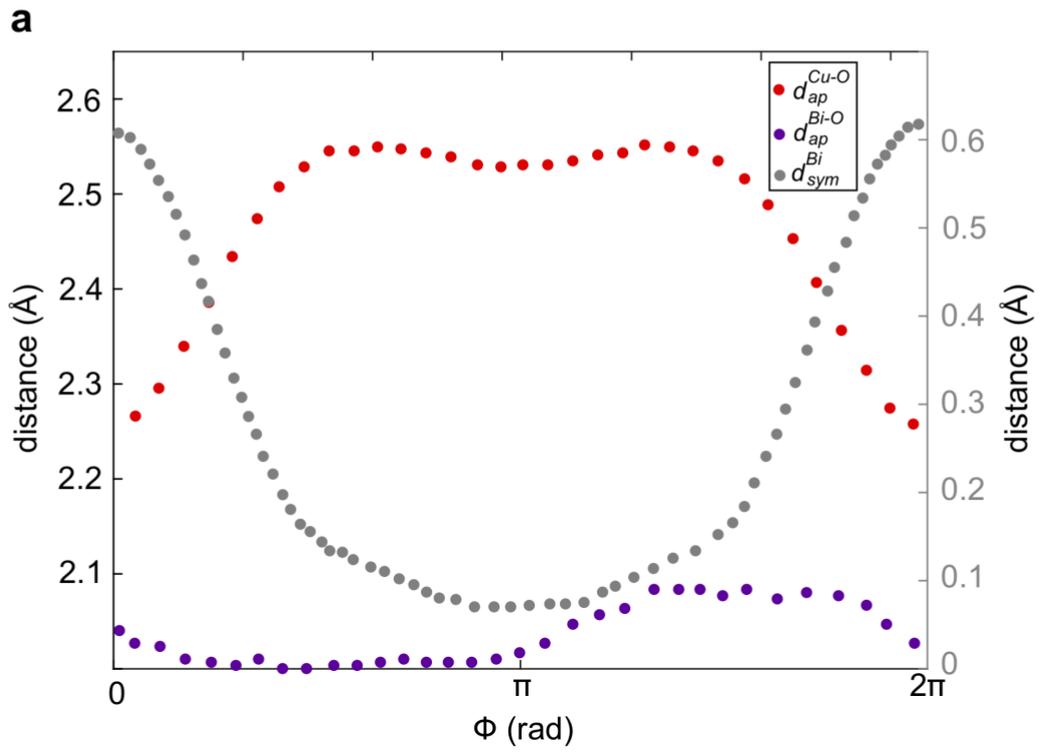

**b**

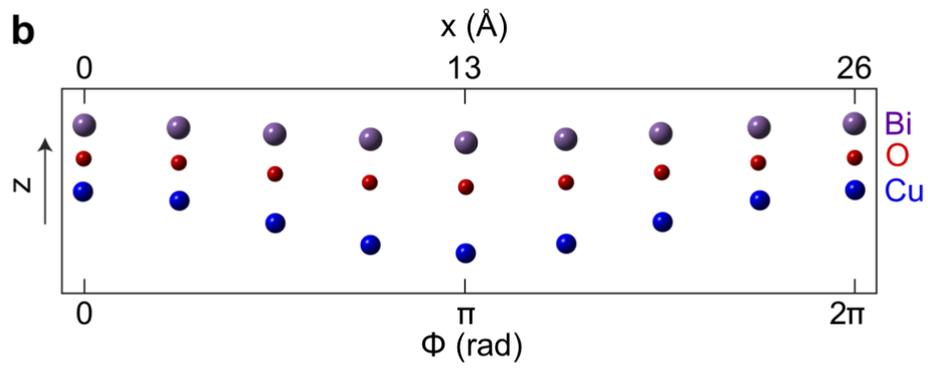

**SM Figure 1: Bi$_2$Sr$_2$CaCu$_2$O$_{8+x}$ Crystal Supermodulation**

a. Left hand axis: apical Cu-O and Bi-O distances as a function of supermodulation phase $\Phi$ from Ref. 53. Right hand axis: displacement of Bi-O termination layer from symmetry plane of crystal as a function of supermodulation phase $\Phi$ from X-ray data. The latter measures the same quantity as the STM topograph $T(\boldsymbol{r})$ enabling one to form a one-to-one mapping between the supermodulation phase $\Phi$ we extract from $T(\boldsymbol{r})$ and the phase variable in the X-ray refinement. Further, the relatively small amplitude of Bi-O modulation compared to the Cu-O apical modulation enables one to make a direct connection between $\Phi$ and $\delta$.

b. Schematic showing the atomic modulations induced along the z axis by the supermodulation. The CuO$_2$ layer modulates with a larger amplitude than the SrO layer, resulting in $\delta$ being maximal at $\Phi = \pi$, where the topograph elevation has its minimum.

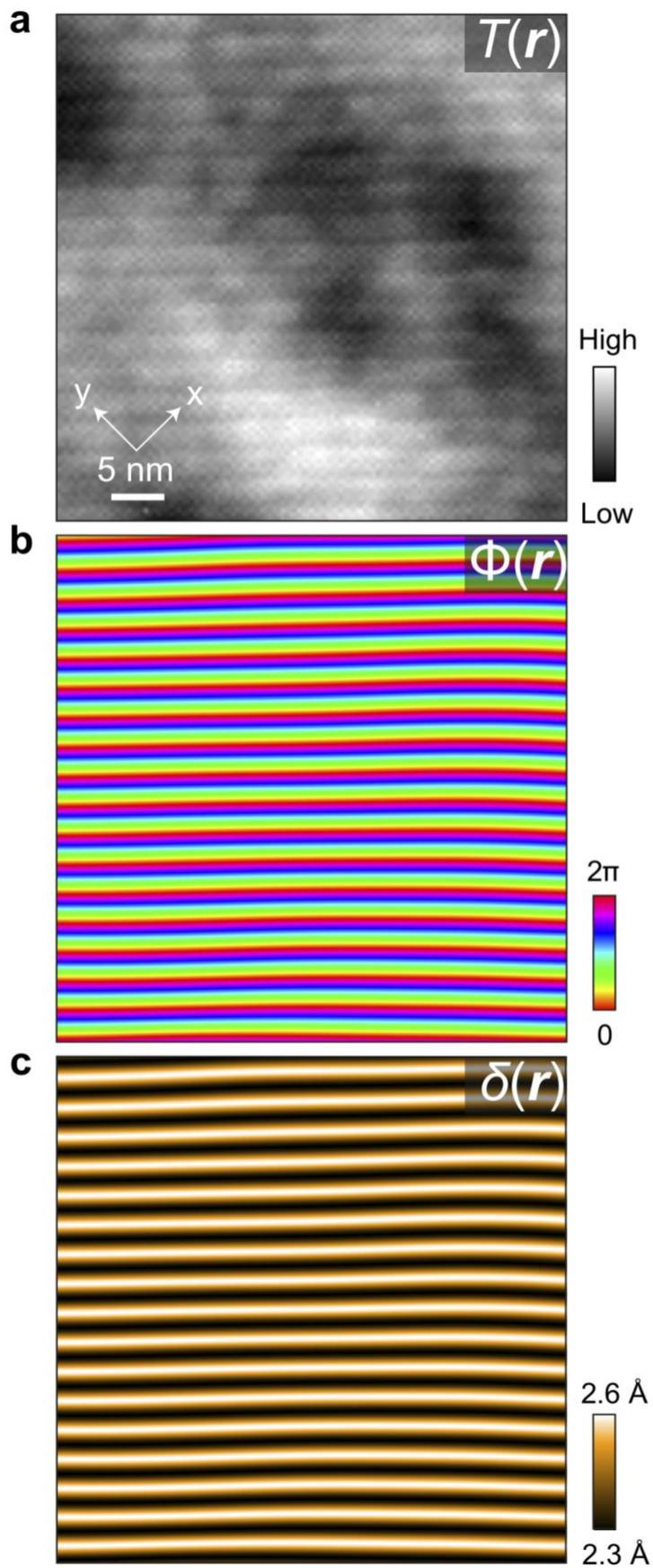

**a** $T(\boldsymbol{r})$

High

Low

y    x

5 nm

**b** $\Phi(\boldsymbol{r})$

$2\pi$

0

**c** $\delta(\boldsymbol{r})$

2.6 Å

2.3 Å

*13*

**SM Figure 2: Bi$_2$Sr$_2$CaCu$_2$O$_{8+x}$ Nanoflake Topography**

Evaluation of a: $T(\boldsymbol{r})$, b: $\varPhi(\boldsymbol{r})$ and c: $\delta(\boldsymbol{r})$ when using the Bi$_2$Sr$_2$CaCu$_2$O$_{8+x}$ nanoflake tip (*7,60*). The atomic structure of the Bi-O termination layer and the supermodulation are clearly visible in $T(\boldsymbol{r})$. The supermodulation phase can be readily extracted from the total $T(\boldsymbol{r})$ signal and is related to $\delta(\boldsymbol{r})$ through the fitting in Eq. S7.

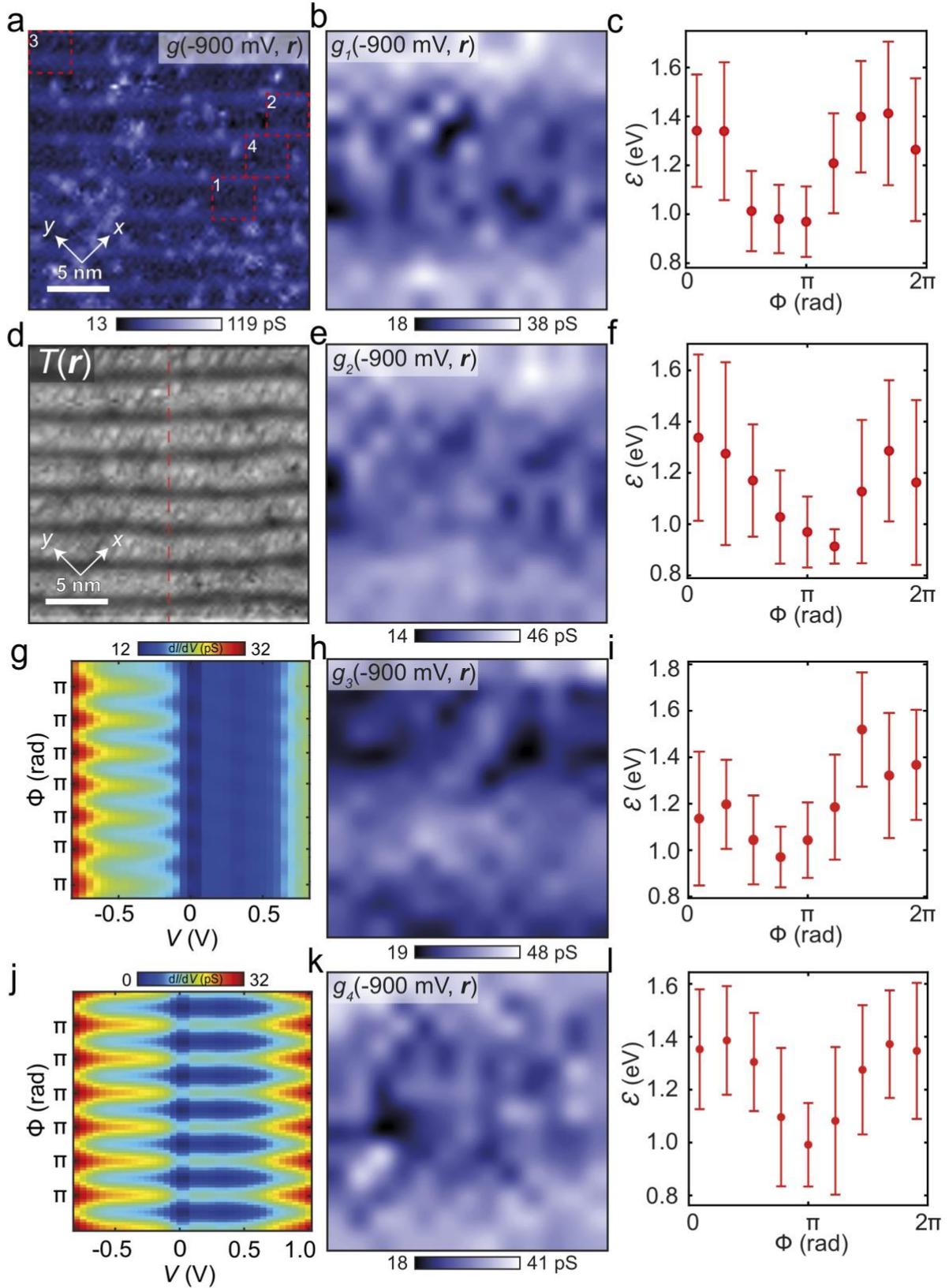

**SM Figure 3: Periodic Modulations in High-Voltage $g(r, V)$ and Low-Voltage $I_m(r, V)$**

a. $g(-900\,\text{mV},\ r)$ showing the distribution of O dopants. Four regions without any dopants have been indicated with dashed red boxes.

b. $g(-900\,\text{mV},\ r)$ in dopant-free region 1.

c. $\mathcal{E}(\Phi)$ obtained in dopant-free region 1.

d. $T(r)$ showing tip-trajectory along a direction perpendicular to the supermodulation as a red dashed line.

e. $g(-900\,\text{mV},\ r)$ in dopant-free region 2.

f. $\mathcal{E}(\Phi)$ obtained in dopant-free region 2.

g. Spectrogram of $g(r, V)$ along the trajectory indicated by the red dashed line in a. Modulations of the lower filled bands are clearly visible. The upper empty states modulate very weakly.

h. $g(-900\,\text{mV},\ r)$ in dopant-free region 3.

i. $\mathcal{E}(\Phi)$ obtained in dopant-free region 3.

j. Simulated spectrogram that would occur due to a setup effect arising from the topograph supermodulation. Both the filled and empty states modulate strongly.

k. $g(-900\,\text{mV},\ r)$ in dopant-free region 4.

l. $\mathcal{E}(\Phi)$ obtained in dopant-free region 4.

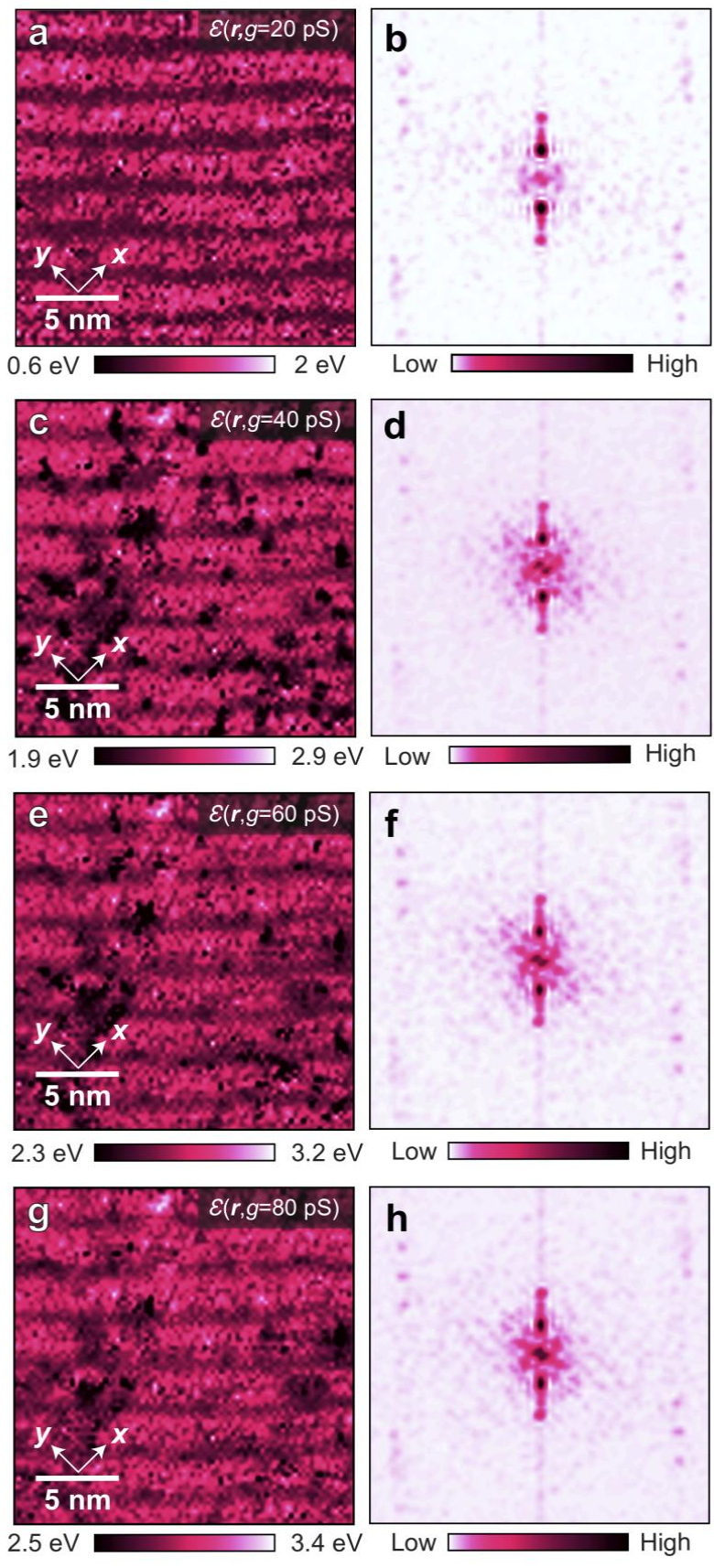

**a** $\mathcal{E}(\boldsymbol{r}, g$=20 pS$)$

**b**

y x

5 nm

0.6 eV ▬▬▬ 2 eV          Low ▬▬▬ High

**c** $\mathcal{E}(\boldsymbol{r}, g$=40 pS$)$

**d**

y x

5 nm

1.9 eV ▬▬▬ 2.9 eV          Low ▬▬▬ High

**e** $\mathcal{E}(\boldsymbol{r}, g$=60 pS$)$

**f**

y x

5 nm

2.3 eV ▬▬▬ 3.2 eV          Low ▬▬▬ High

**g** $\mathcal{E}(\boldsymbol{r}, g$=80 pS$)$

**h**

y x

5 nm

2.5 eV ▬▬▬ 3.4 eV          Low ▬▬▬ High

**SM Figure 4: Imaging $\mathcal{E}(G, \boldsymbol{r})$ in Different $G$**

Calculated $\mathcal{E}(G, \boldsymbol{r})$ for input differential conductance $G$ of a: $20\,\text{pS}$, c: $40\,\text{pS}$, e: $60\,\text{pS}$ and g: $80\,\text{pS}$ showing that the spatial structure of $\mathcal{E}$ is relatively insensitive to our choice of $G$, apart from the presence of Oxygen dopant peaks at $G \geq 40\,\text{pS}$. Panels b, d, f and h show the corresponding power spectral densities, which have clear peaks at $\boldsymbol{q} \approx \pm \boldsymbol{Q}_S$ for all values $G$ considered.

**a**

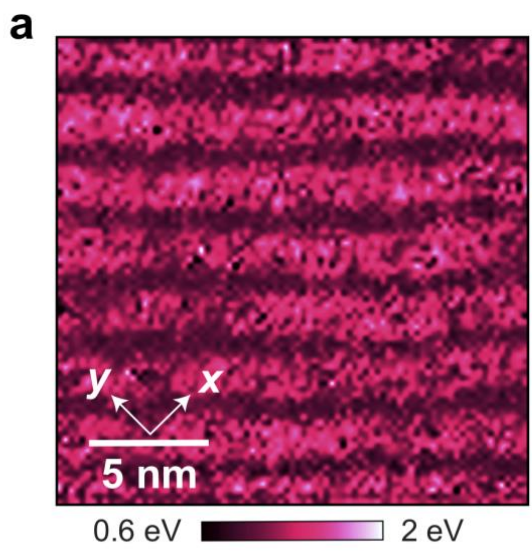

y  x

5 nm

0.6 eV ▬▬▬▬ 2 eV

**b**

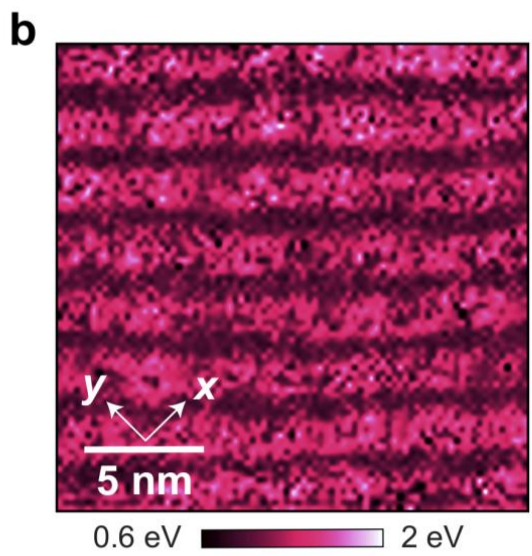

y  x

5 nm

0.6 eV ▬▬▬▬ 2 eV

**c**

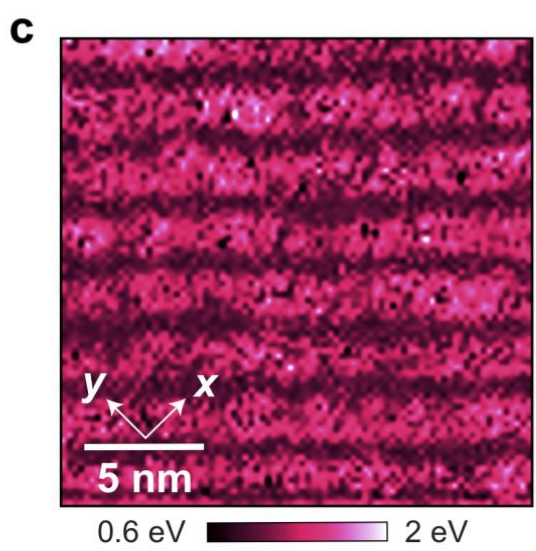

y  x

5 nm

0.6 eV ▬▬▬▬ 2 eV

**d**

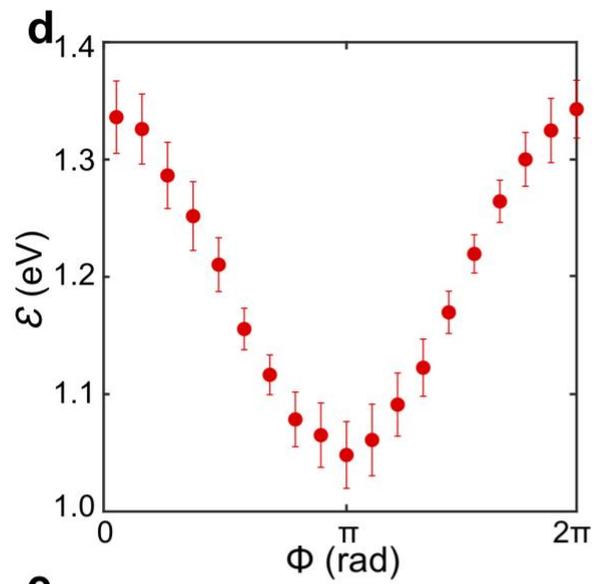

**e**

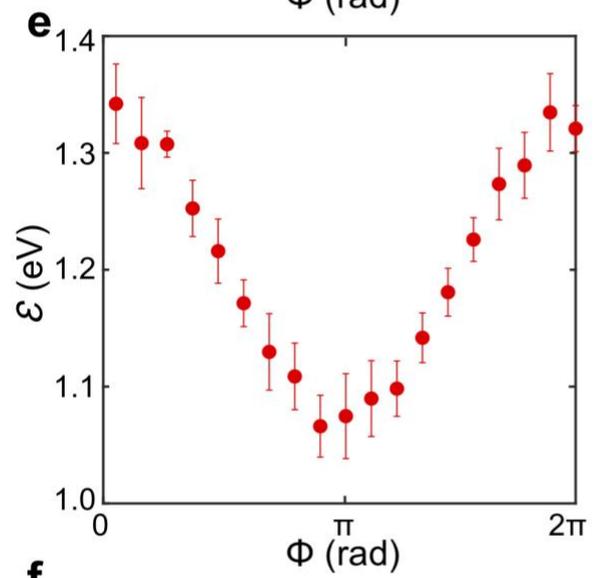

**f**

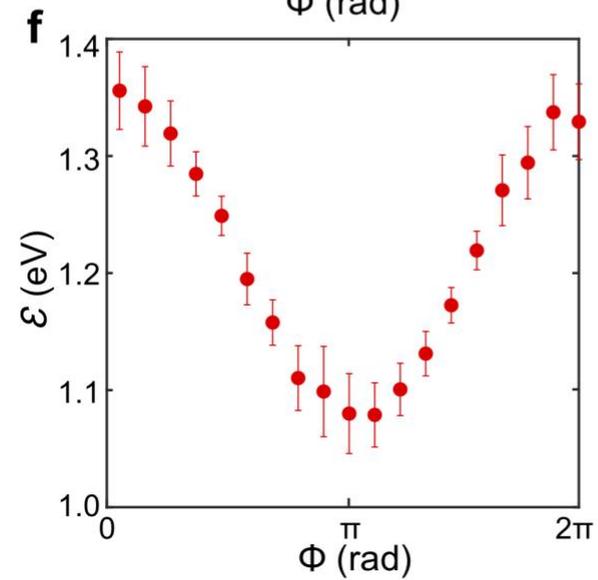

**SM Figure 5: Imaging $\mathcal{E}(\boldsymbol{r})$ in Different Fields of View**.

(a-c) $\mathcal{E}(\boldsymbol{r})$ calculated in 3 disjoint $19.5 \times 19.5 \text{ nm}^2$ fields of view.

(d-f) Phase-averaged $\tilde{\mathcal{E}}(\varPhi)$ calculated in 3 disjoint $19.5 \times 19.5 \text{ nm}^2$ fields of view. $\tilde{\mathcal{E}}(\varPhi)$ is peaked at $\varPhi \approx \pi$ with a peak-to-peak amplitude of $\sim 0.3$ eV in all three fields of view.

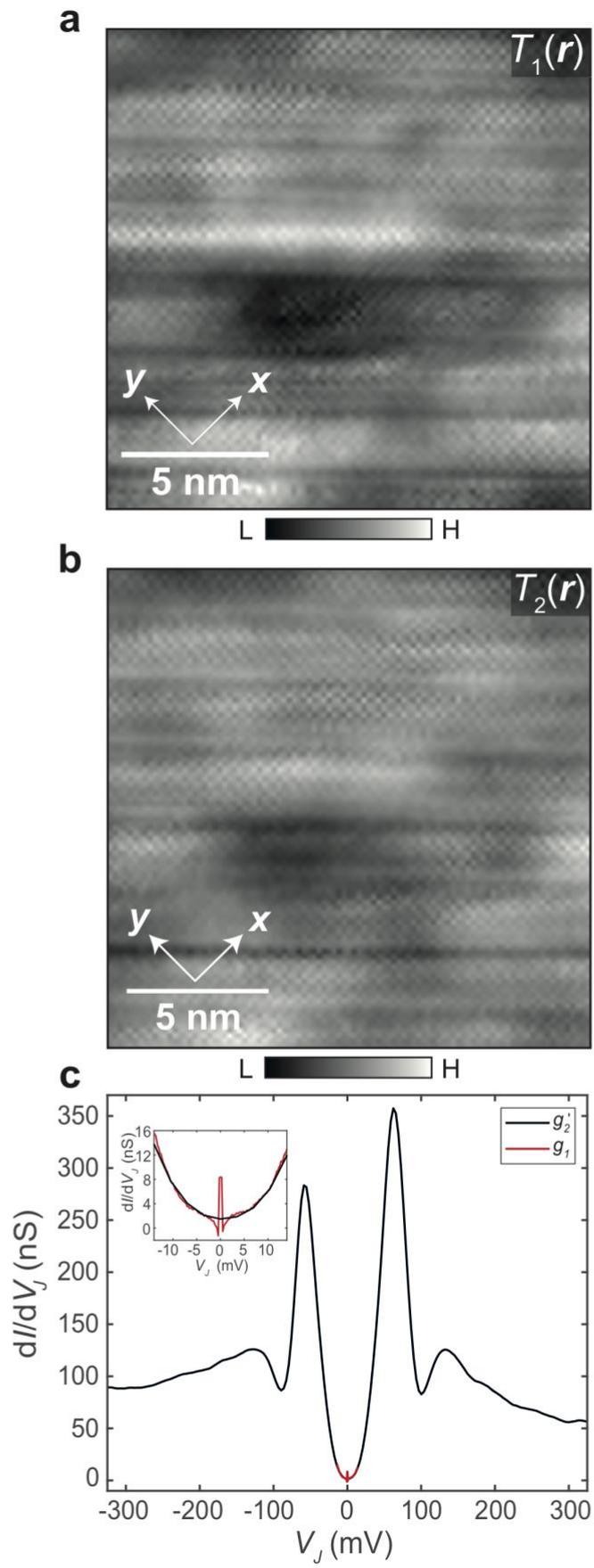

**SM Figure 6: Scaling High-Voltage and Low-Voltage $g(r, V_J)$ for $R_N(r)$**

(a-b) Simultaneous topographies of $g_1(r, V_J)$ and $g_2(r, V_J)$.

c. Typical example of a rescaled $g_2'(V_J)$ and the original $g_1(V_J)$ spectra. Inset: Same spectra plotted for voltages in the range $-V_{s1} \leq V_J \leq V_{s1}$.

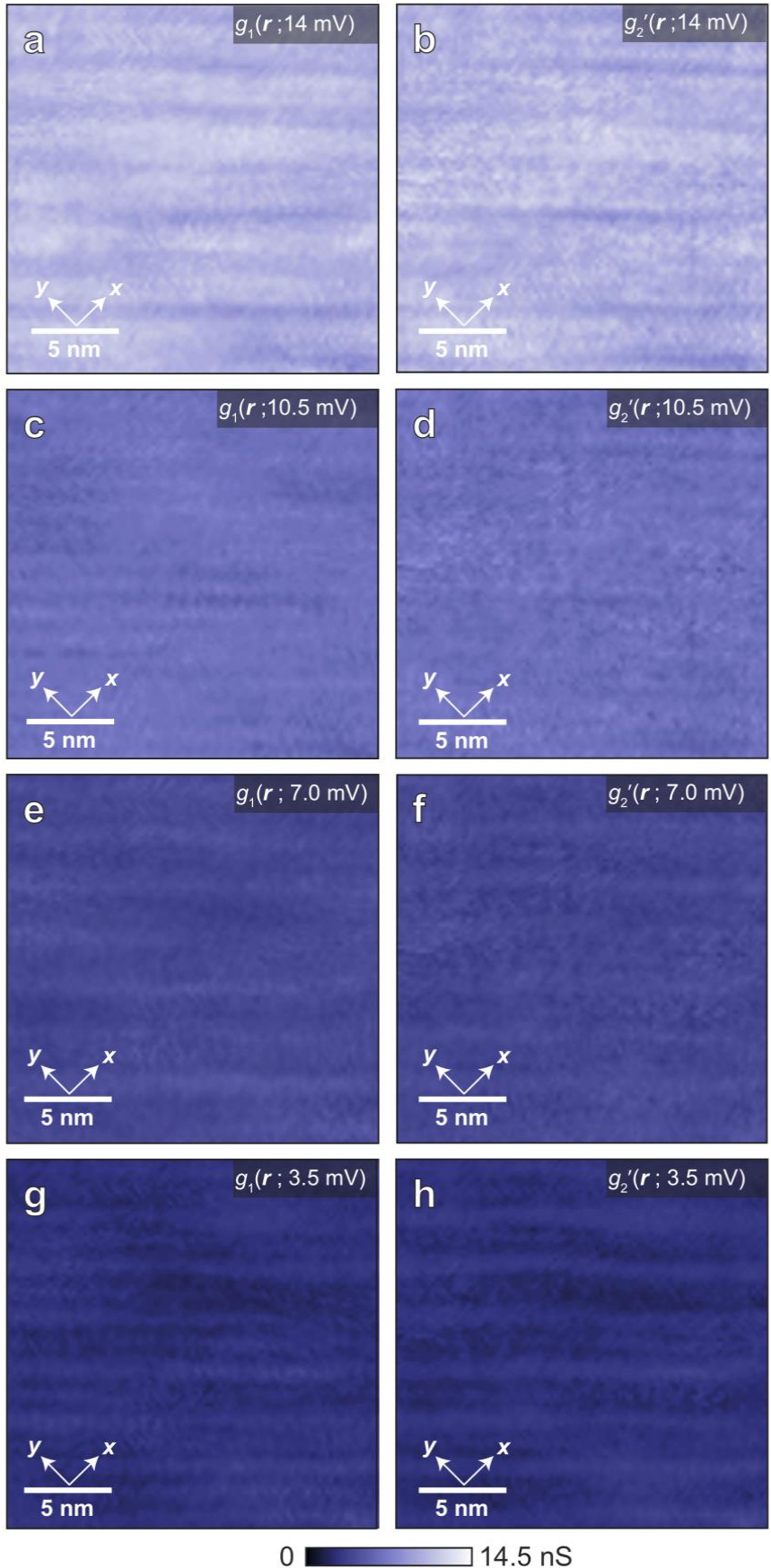

$g_1(\boldsymbol{r};14\,\text{mV})$    $g_2'(\boldsymbol{r};14\,\text{mV})$

$g_1(\boldsymbol{r};10.5\,\text{mV})$    $g_2'(\boldsymbol{r};10.5\,\text{mV})$

$g_1(\boldsymbol{r};7.0\,\text{mV})$    $g_2'(\boldsymbol{r};7.0\,\text{mV})$

$g_1(\boldsymbol{r};3.5\,\text{mV})$    $g_2'(\boldsymbol{r};3.5\,\text{mV})$

0      14.5 nS

**SM Figure 7: Rescaled $g(r, V)$ Images**

(a-h) The rescaled $g'_2(r, V)$ and original $g_1(r, V)$ images compared at voltages ranging from $3.5$ mV to $14$ mV. They are virtually identical, validating the scaling procedure used to obtain $R_N(r)$.

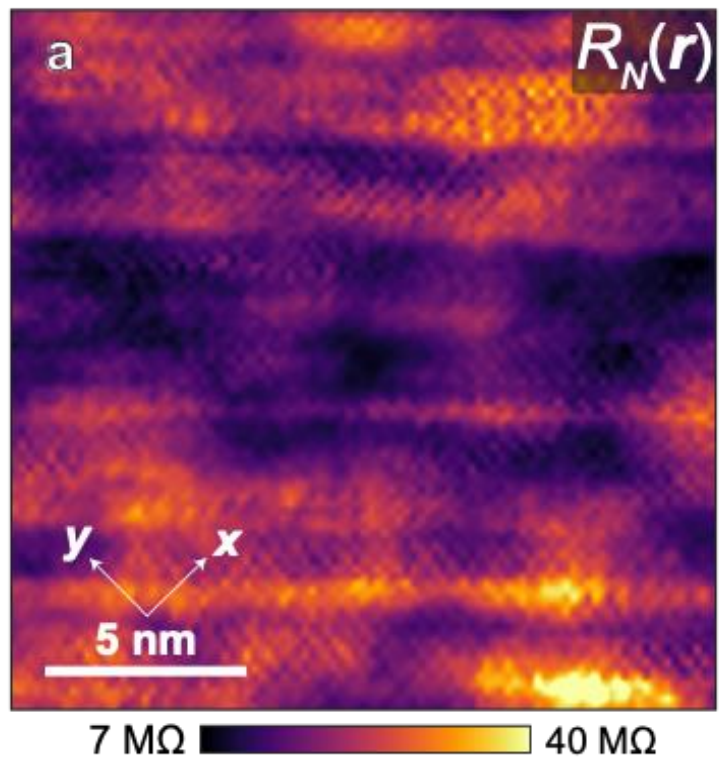

a  $R_N(\boldsymbol{r})$

y  x

5 nm

7 MΩ ▮▮▮▮▮▮▮▮ 40 MΩ

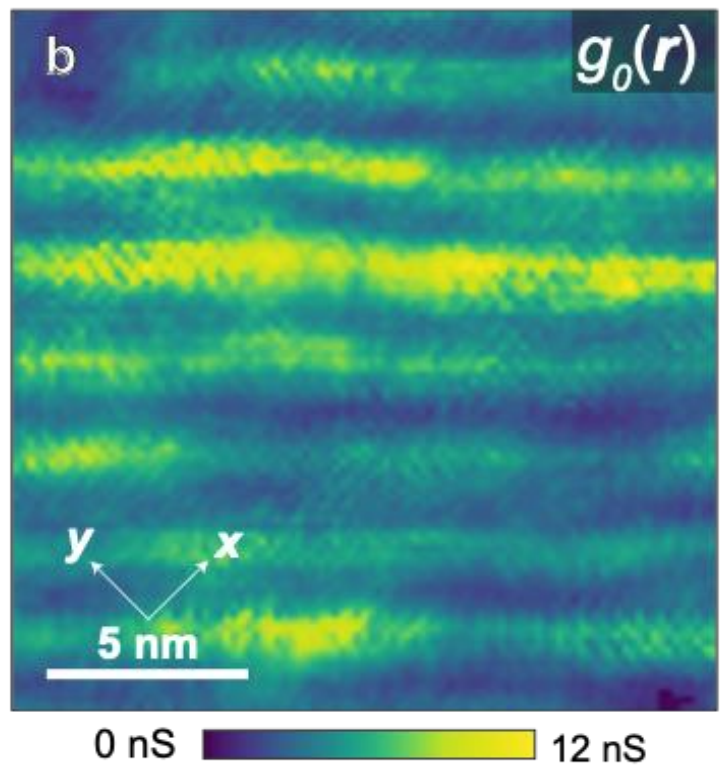

b  $g_0(\boldsymbol{r})$

y  x

5 nm

0 nS ▮▮▮▮▮▮▮▮ 12 nS

**SM Figure 8: Imaging $R_N(r)$ and $g_0(r)$**

(a) Image of the normal-state junction resistance $R_N(r) \equiv 1/g_2'(r, V_{s2})$ at which the $g_1(r, V_J)$ map was measured.

(b) Image of $g_0(r)$ in the same field of view.

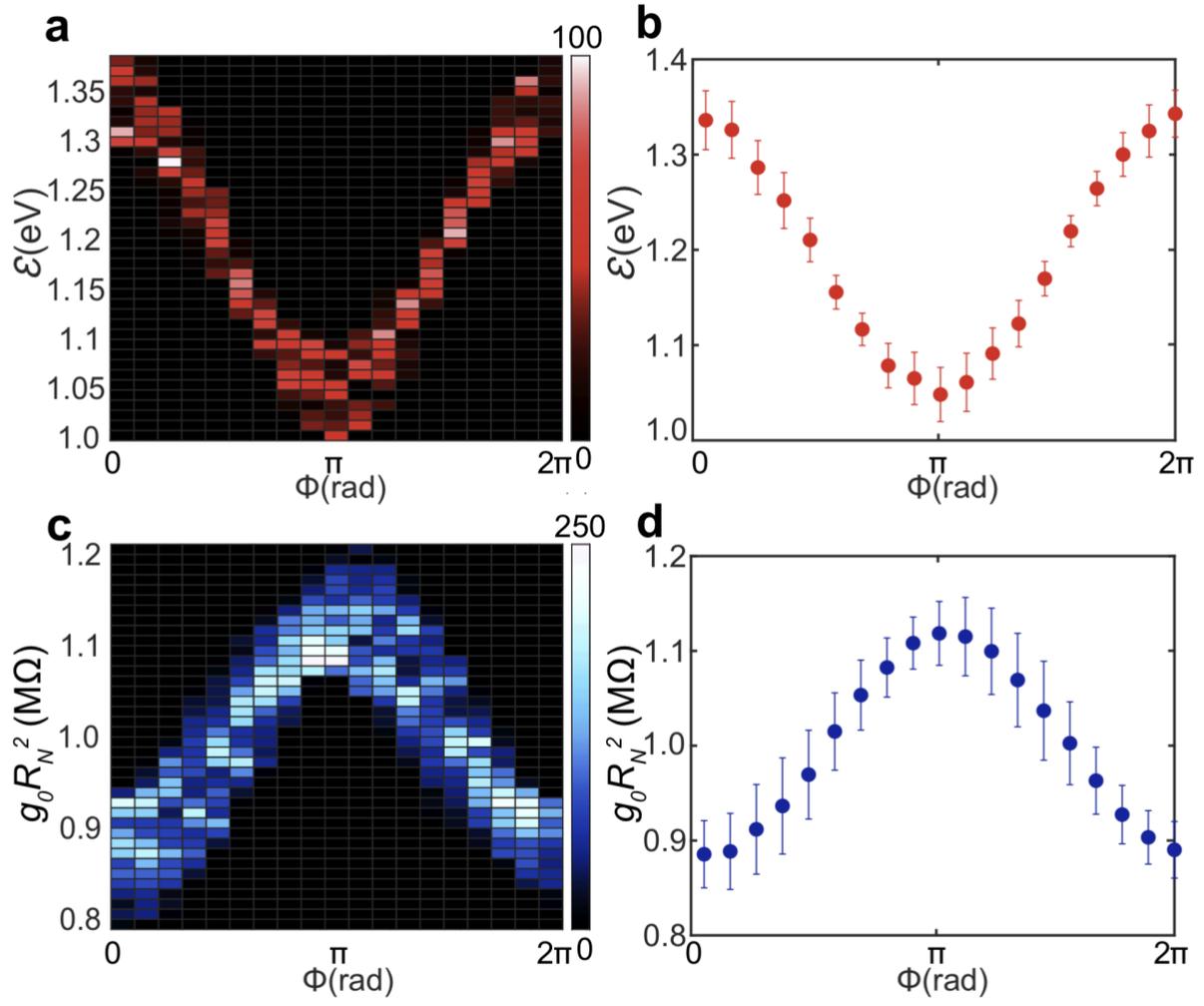

**SM Figure 9: Determining $\tilde{\mathcal{E}}(\Phi)$ and $\tilde{n}_P(\Phi)$**

a. Two-dimensional histogram of $\tilde{\mathcal{E}}$ vs $\Phi$, clearly showing a minimum at $\Phi \approx \pi$.

b. $\tilde{\mathcal{E}}(\Phi)$ obtained by averaging $\tilde{\mathcal{E}}$ within each phase bin.

c. Two-dimensional histogram of $\tilde{n}_P \propto g_0(r)R_N{}^2(r)$ vs $\Phi$, clearly showing a maximum at $\Phi \approx \pi$.

d. $\tilde{n}_P(\Phi) \propto g_0 R_N{}^2(\Phi)$ obtained by averaging $\tilde{n}_P$ within each phase bin.

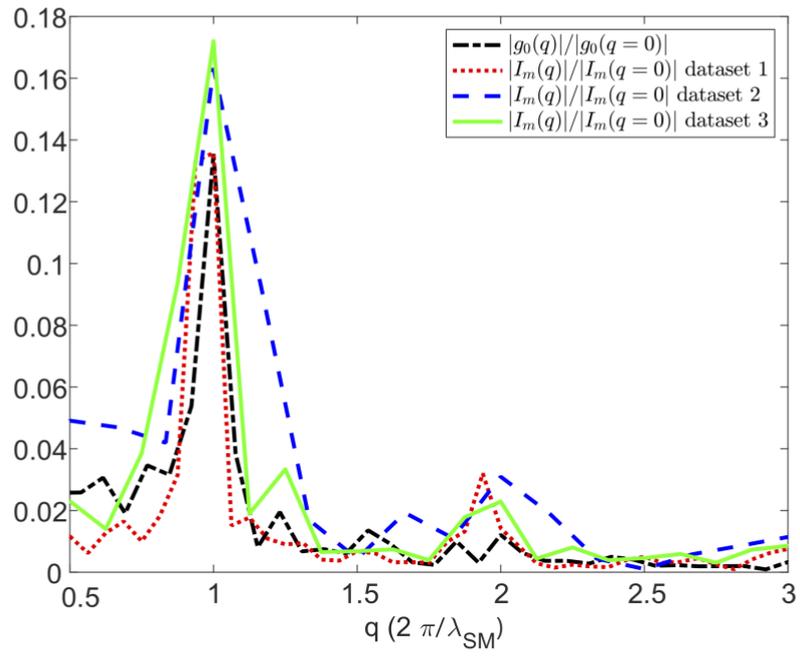

**SM Figure 10: Repeatability of $I_m$ Supermodulation Amplitude**

Fourier transform linecuts of $I_m$ from three different experiments in three different fields of view along with the equivalent linecut for $g_0$. The black dashed curve corresponds to the $g_0$ data used to compute electron-pair density $n_p$ in the main text.